\documentclass[english]{article}
\usepackage[T1]{fontenc}
\usepackage[latin1]{inputenc}
\usepackage{geometry}
\usepackage{subfigure}
\usepackage{amsmath}
\usepackage{graphicx}
\usepackage{amssymb}
\IfFileExists{url.sty}{\usepackage{url}}
                      {\newcommand{\url}{\texttt}}

\makeatletter

\providecommand{\tabularnewline}{\\}

 \newcommand{\lyxaddress}[1]{
   \par {\raggedright #1
   \vspace{1.4em}
   \noindent\par}
 }

\date{}

\usepackage{babel}
\makeatother
\begin{document}

\title{Spontaneous polarization in eukaryotic gradient sensing: A mathematical
model based on mutual inhibition of frontness and backness pathways}

\author{Atul Narang%
\footnote{Email: \texttt{narang@che.ufl.edu}, URL: \protect\url{http://narang.che.ufl.edu}%
}}

\maketitle

\lyxaddress{Department of Chemical Engineering, University of Florida, Gainesville,
FL 32611-6005}

\begin{abstract}
A key problem of eukaryotic cell motility is the signaling mechanism
of chemoattractant gradient sensing. Recent experiments have revealed
the molecular correlate of gradient sensing: Frontness molecules,
such as PI3P and Rac, localize at the front end of the cell, and backness
molecules, such as Rho and myosin~II, accumulate at the back of the
cell. Importantly, this frontness-backness polarization occurs {}``spontaneously''
even if the cells are exposed to uniform chemoattractant profiles.
The spontaneous polarization suggests that the gradient sensing machinery
undergoes a Turing bifurcation. This has led to several classical
activator-inhibitor and activator-substrate models which identify
the frontness molecules with the activator. Conspicuously absent from
these models is any accounting of the backness molecules. This stands
in sharp contrast to experiments which show that the backness pathways
inhibit the frontness pathways. Here, we formulate a model based on
the mutually inhibitory interaction between the frontness and backness
pathways. The model builds upon the mutual inhibition model proposed
by Bourne and coworkers (Xu \emph{et al,} \emph{Cell,} \textbf{114},
201--214, 2003). We show that mutual inhibition alone, without the
help of any positive feedback, can trigger spontaneous polarization
of the frontness and backness pathways. The spatial distribution of
the frontness and backness molecules in response to inhbition and
activation of the frontness and backness pathways are consistent with
those observed in experiments. Furthermore, depending on the parameter
values, the model yields spatial distributions corresponding to chemoattraction
(frontness pathways in-phase with the external gradient) and chemorepulsion
(frontness pathways out-of-phase with the external gradient). Analysis
of the model suggests a mechanism for the chemorepulsion-to-chemoattraction
transition observed in neurons.
\end{abstract}

\section{\label{s:Introduction}Introduction}

\begin{figure}
\begin{center}\subfigure[]{\includegraphics[%
  clip,
  width=3in,
  height=2in]{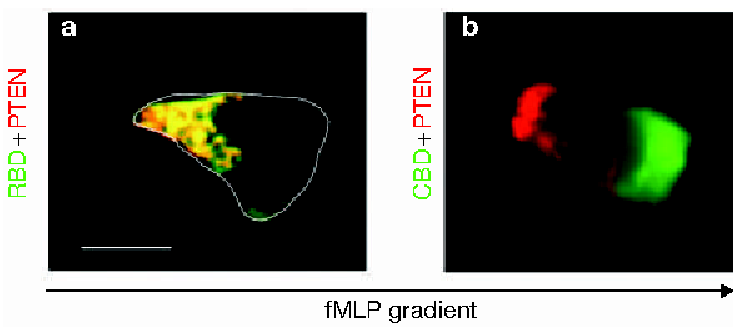}}\hspace{0.1in}\subfigure[]{\includegraphics[%
  clip,
  width=2in,
  height=2in]{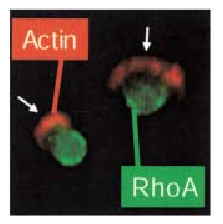}}\end{center}

\caption{\label{f:SpatialSegregation}Spatial segregation of frontness and
backness molecules. (a) When neutrophils are exposed to a gradient
of fMLP, Rho (represented by its marker,~RBD) and PTEN colocalize
at the back of the cells, and Cdc42 (represented by its marker,~CBD)
localizes to the front of the cells (from~\cite{Li2005}). (b)~The
spatial segregation occurs even in the absence of a chemoattractant
gradient. When neutrophil-like HL-60 cells are exposed to a uniform
concentration (100~nM) of fMLP, actin polymers (shown in red) localize
at the front of the cell, and Rho (shown in green) localizes at the
back of the cell (from~\cite{xu03b}).}
\end{figure}

When motile cells are exposed to a chemoattractant gradient, they
develop a morphological polarity consisting of a distinct front and
back~\cite{lauffenburger96}. The formation of the morphological
polarity is driven by the spatial segregation of distinct sets of
intracellular molecules to the front and the rear of the cell (see
Figure~\ref{f:SpatialSegregation}a and refs.~\cite{funamoto02,iijima02a,xu03b,Li2005}).
The \emph{frontness} molecules, which include Cdc42, Rac, PI3P, PI3K,
Arp2/3, and F-actin, localize to the front of the cell where they
coordinate the extension of an actin-rich extension. The \emph{backness}
molecules, which include Rho, Rho kinase, PTEN, and myosin~II, migrate
to the rear of the cell where they are thought to activate cell contraction.

In general, the spatial segregation of the frontness/backness molecules
and the resultant morphological polarization occurs even if the cells
are exposed to a uniform chemoattractant profile~(see Figure~\ref{f:SpatialSegregation}b
and \cite{wang02,weiner02b,postma03}). This phenomenon has been called
\emph{spontaneous polarization} to emphasize the fact that the cells
polarize despite the absence of a perceptible external cue~\cite{wedlich-soldner03a}.

The existence of spontaneous polarization is reminiscent of the Turing
instability in reaction-diffusion systems~\cite{turing52}. Thus,
it has led to several models that view spontaneous polarization, either
explicitly or implicitly, as the onset of a Turing bifurcation~\cite{meinhardt99,narang01,postma01,Maly2004,Subramanian2004a}.%
\footnote{Alternative models of gradient sensing in which spontaneous polarization
plays no role have also been proposed~\cite{Levchenko02,rappel02,krishnan03,Ma2004,schneider04}.
For the most part, they are motivated by experimental systems such
as PDGF-stimulated fibroblasts~\cite{schneider04} and latrunculin-treated
\emph{Dictyostelium} cells~\cite{Levchenko02}, which do not display
spontaneous polarization. In these models, polarization occurs only
in the presence of external gradients. %
} According to these models, the cell is in a stable homogeneous steady
state in the absence of chemoattractant. However, exposure of the
cell to a sufficiently large uniform chemoattractant concentration
pushes it past a Turing bifurcation point, where the homogeneous steady
state is unstable with respect to certain non-homogeneous perturbations.
The inevitable presence of noise is then sufficient to drive the cell
towards a nonhomogeneous steady state corresponding to the polarized
state of the cell.

\begin{figure}
\begin{center}\subfigure[]{\includegraphics[%
  clip,
  width=3in,
  height=2in]{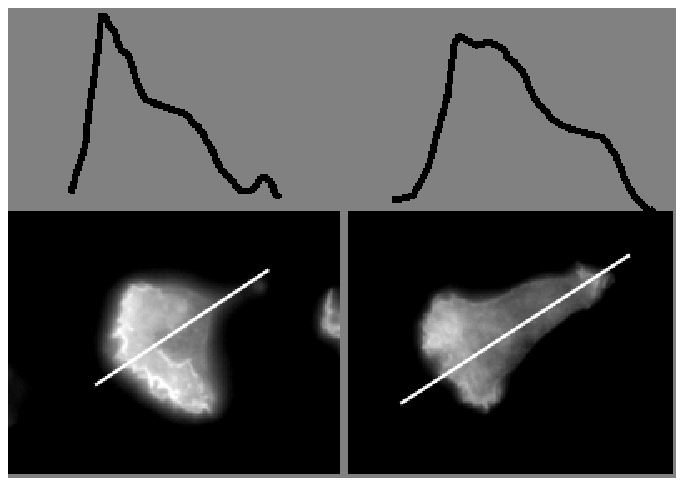}}\hspace{0.3in}\subfigure[]{\includegraphics[%
  clip,
  width=1in,
  height=2in]{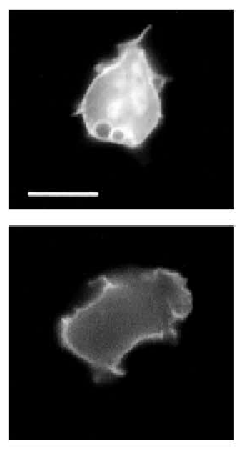}}\end{center}

\caption{\label{f:FrontnessInhibitionActivation}Spatial redistribution of
the frontness pathways in response to inhibition and activation of
the frontness pathways. (a) When HL-60 cells are pretreated with intermediate
concentrations (100--200~$\mu$M) of LY294002, an inhibitor of PI3K,
before their exposure to fMLP, the gradient of PH-Akt, a marker for
PI3P, decreases. The figures on the left and right show the spatial
distribution and fluorescence intensity of PH-Akt in control and LY294002-treated
cells, respectively. At high concentrations (300~$\mu$M) of LY294002,
PH-Akt does not polarize at all (from~\cite{wang02}). (b)~When
HL-60 cells transfected with constitutively active Rac are exposed
to fMLP, the frontness molecules, PI3P (top) and active Rac (bottom),
spread all over the membrane (from~\cite{srinivasan03}).}
\end{figure}

\begin{figure}
\begin{center}\subfigure[]{\includegraphics[%
  clip,
  width=2in,
  height=3in]{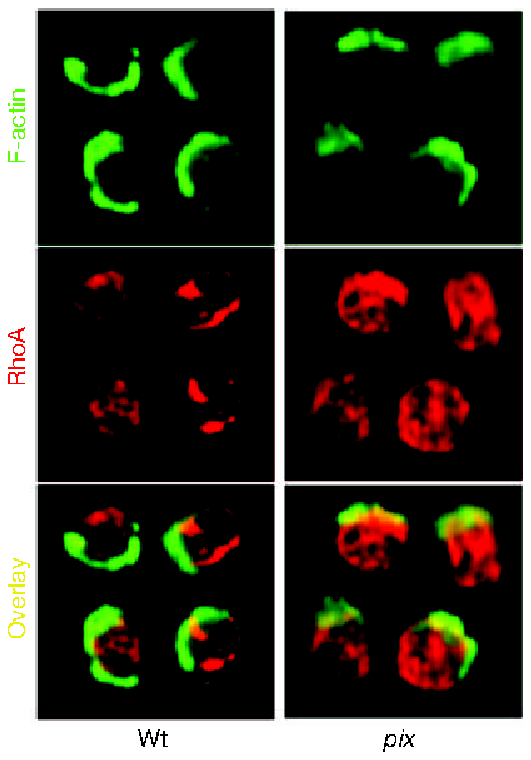}}\hspace{0.1in}\subfigure[]{\includegraphics[%
  clip,
  width=2in,
  height=3in]{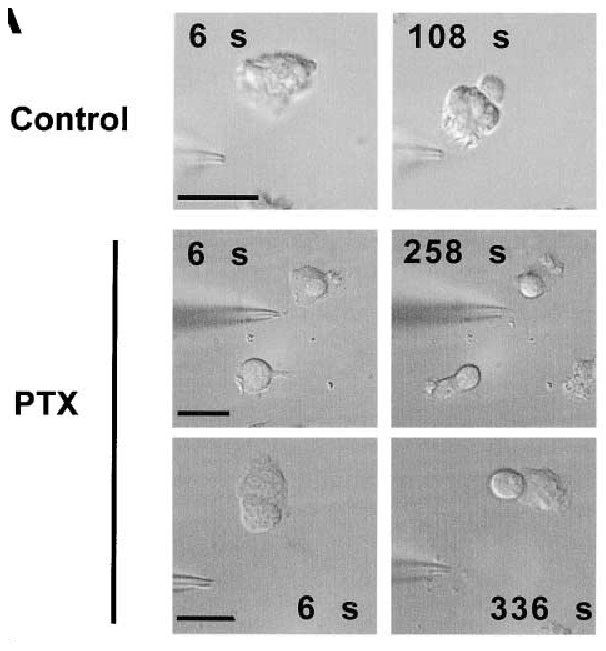}}\end{center}

\caption{\label{f:Chemorepulsion}Spatial redistribution of the backness pathways
in response to inhibition of the frontness pathways. (a)~When neutrophils
devoid of $\alpha$-Pix, an activator of Cdc42, polarize spontaneously,
the backness component RhoA is found at the back as well as the front
of the cell (along with actin polymers). The panels on the left and
right show the distribution of F-actin~(green), RhoA~(red) and the
overlay in wild-type and $\alpha$-Pix-null cells (from~\cite{Li2005}).
(b)~HL-60 cells treated with PTX, a potent inhibitor of Gi proteins,
extend a uropod-like structure at the edge exposed to the chemoattractant
source (from~\cite{xu03b}). In control cells (top panel), a broad
{}``head'' is extended toward the fMLP source. In PTX-treated cells
(middle and bottom panel), a knob-like structure, similar to the {}``tail''
of control cells, is extended toward the fMLP source.}
\end{figure}

\begin{figure}
\begin{center}\subfigure[]{\includegraphics[%
  clip,
  width=3in,
  height=2in]{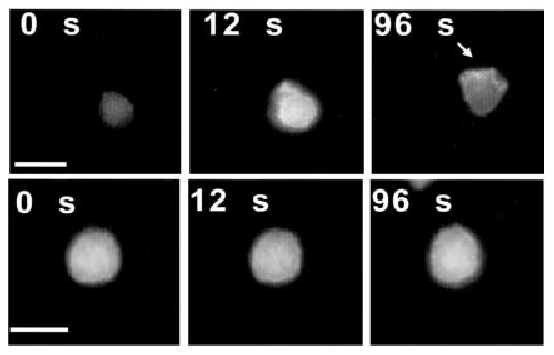}}\hspace{0.3in}\subfigure[]{\includegraphics[%
  clip,
  width=3in,
  height=1in]{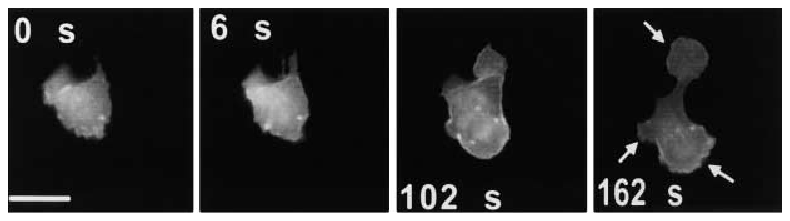}}\end{center}

\caption{\label{f:BacknessInhibitionActivation}Spatial redistribution of
the frontness and backness pathways in response to inhibition and
activation of the backness pathways (from~\cite{xu03b}). (a)~When
HL-60 cells transfected with constitutively active Rho are exposed
to fMLP, there is no polarization of PI3P. The top and lower panel
show the evolution of the spatial distribution of PI3P in wild-type
and transfected cells. (b)~When HL-60 cells transfected with dominant
negative Rho are exposed to a uniform concentration of fMLP, PH-Akt,
a marker for PI3P, spreads all over the membrane, and multiple pseudopods
develop. }
\end{figure}

As attractive as these models of spontaneous polarization may be,
there is a significant gap between the theory and experiments. Indeed,
all the models consist of one or more slow-diffusing \emph{activators}
whose synthesis is autocatalytic, and which in consequence, tend to
grow and spread across the entire cell. The unconstrained growth and
dispersion of the activator is restricted by hypothesizing the existence
of a diffusible \emph{inhibitor} (which is a by-product of activator
synthesis and impedes the growth of the activator) or \emph{substrate}
(which is consumed during activator synthesis and stimulates the growth
of the activator). The diffusible activator/substrate ensures that
the growth of the activator(s) remains confined to a localized region
of the cell membrane. This region is identified with the front of
the cell, and the rest of the cell membrane, suffering from an activator
deficit, is presumed to constitute the back of the cell. The predictions
of these models are partially consistent with experiments involving
the activation or inhibition of the frontness molecules. Indeed, if
PI3K is inhibited, the gradient of the frontness molecules such as
PI3P progressively decreases, until at sufficiently high levels of
inhibition, there is no gradient at all ~(Figure~\ref{f:FrontnessInhibitionActivation}a).
On the other hand, when Rac is overexpressed, high levels of frontness
molecules, PI3P and Rac, are found all over the cell membrane~(Figure~\ref{f:FrontnessInhibitionActivation}b).
Both these features are reproduced by the activator-inhibitor class
of models~\cite{narang01,Maly2004,Subramanian2004a}. However, these
models cannot explain the spatial distribution of the backness molecules
in response to inhibition of the frontness molecules. Specifically,
if the activity of Cdc42 is suppressed, Rho is found not only at the
back but also at the front of the cell (Figure~\ref{f:Chemorepulsion}a).
Likewise, if the Gi proteins are inhibited with pertussis toxin (PTX),
a uropod-like structure forms at the up-gradient edge of the cell~(Figure~\ref{f:Chemorepulsion}b).
The models also offer no insight into experiments involving activation
or inhibition of the backness molecules. For instance, when Rho is
activated, PI3P fails to polarize (Figure~\ref{f:BacknessInhibitionActivation}a).
Conversely, when Rho is inhibited, PI3P spreads all over the membrane,
and the cell extends a single broad pseudopod or multiple pseudopods~(Figure~\ref{f:BacknessInhibitionActivation}b).
These results are beyond the scope of the activator-inhibitor and
activator-substrate models because the backness molecules are not
even acknowledged as legitimate variables --- they are implicitly
assumed to somehow settle down in regions uninhabited by the frontness
molecules. This stands in sharp contrast to the experimental data
which shows that backness pathways downregulate the frontness pathways:
Inhibition of Rho kinase increases Rac activity 2--3 fold~\cite{xu03b}.

\begin{figure}
\begin{center}\subfigure[]{\includegraphics[%
  width=3in,
  height=2in]{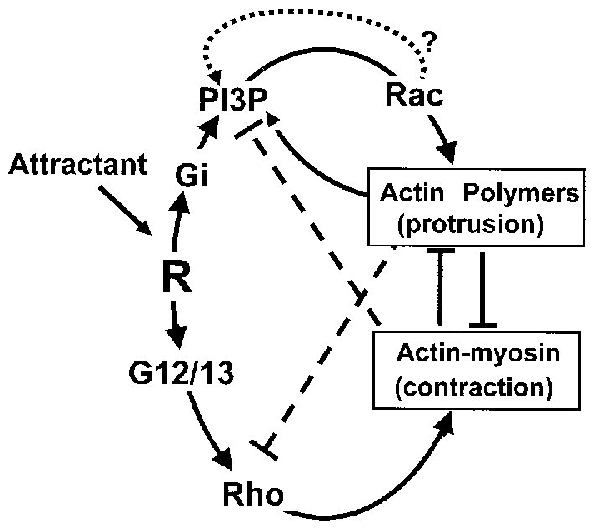}}\hspace{0.5in}\subfigure[]{\includegraphics[%
  width=1.8in,
  height=2in]{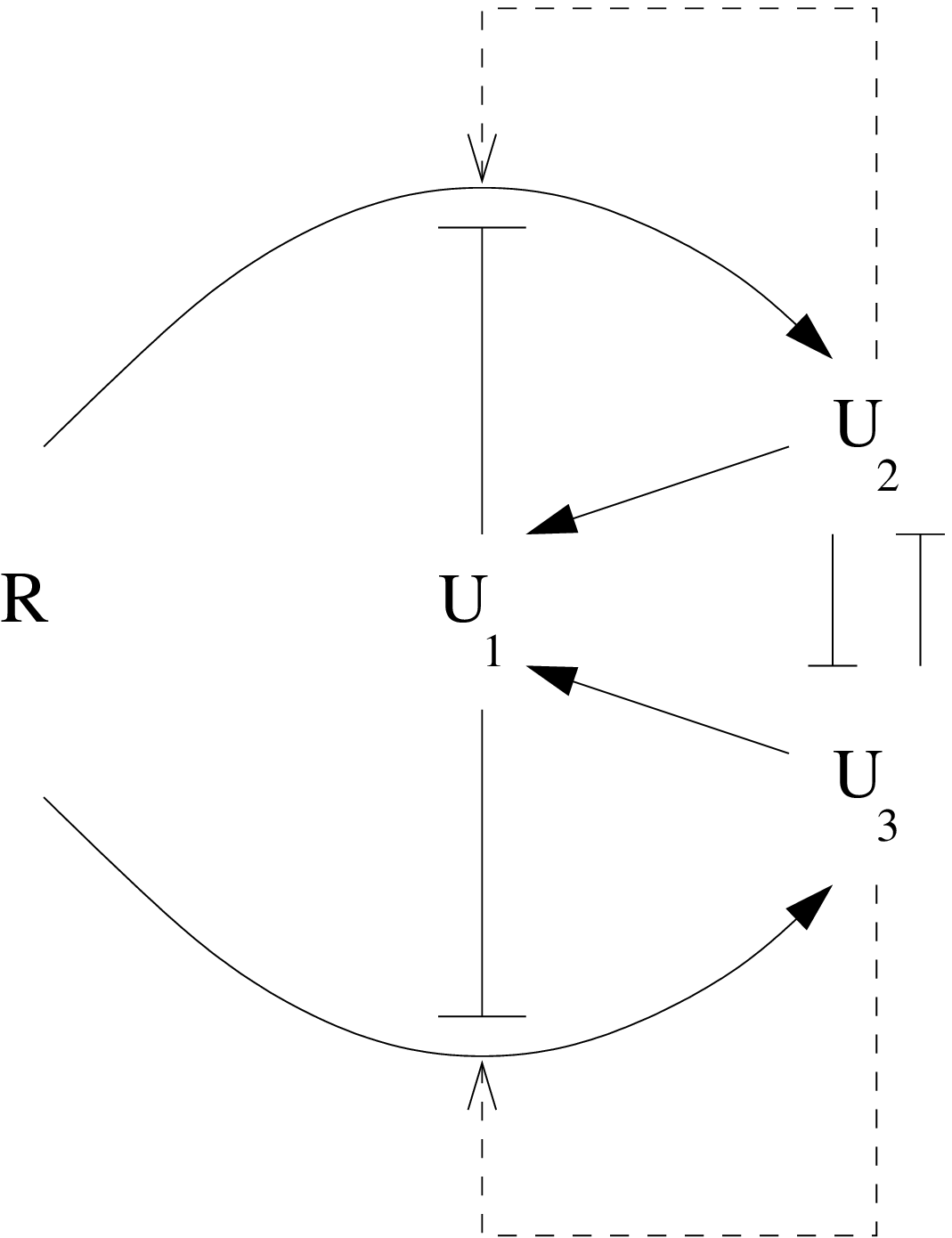}}\end{center}

\caption{\label{f:ModelScheme}Model schemes. (a)~Model proposed by Bourne
and co-workers for spontaneous polarization of neutrophils and HL-60
cells (from~\cite{xu03b}). The chemoattractant binds to G-protein-coupled
receptors, R. The Gi-coupled receptors activate the frontness pathway
which includes PI3P, Rac, and actin polymers. The G12/13-coupled receptors
activate the backness pathway which includes Rho and myosin. The frontness
pathway is subject to positive feedback from actin polymers (full
line from actin polymers to PI3P) and possibly Rac (dashed line from
Rac to PI3P). The frontness and backness pathways inhibit each other
(dashed lines with a bar at the end). (b)~The model scheme considered
in this paper. Receptor activation stimulates the synthesis of frontness
and backness subsystems, denoted $U_{2}$ and $U_{3}$, respectively.
The frontness and backness subsystems stimulate their own synthesis
(dashed lines), but inhibit each other both directly (line from $U_{2}$
to $U_{3}$, and vice versa), and indirectly through production of
the inhibitor, $U_{1}$.}
\end{figure}

One hypothesis regarding the interaction between the frontness and
backness pathways is that they inhibit each other~\cite{xu03b}.
Based on extensive experiments with neutrophils and neutrophil-like
HL-60 cells~\cite{servant99,wang02,weiner02b,srinivasan03,xu03b},
Bourne and coworkers arrived at the kinetic scheme shown in Figure~\ref{f:ModelScheme}a,
which they describe as follows

\begin{quote}
Briefly, the attractant binds to a G protein-coupled receptor (R),
which in turn activates different trimeric G proteins to generate
two divergent, opposing signaling pathways, which promote polarized
frontness and backness, respectively. In the frontness pathway, Gi,
PI3Ps, and Rac promote de novo formation of actin polymers. One or
more positive feedback loops in this first pathway mediate localized
increases in sensitivity to attractant: one of these requires polymerized
actin, while Rac or Cdc 42 may in addition enhance PI3P accumulation
more directly, via an alternative pathway (dotted curved line in Figure~7
{[}reproduced here as Figure~\ref{f:ModelScheme}a{]}). Backness
signals, generated by G12 and G13, depend on activation of a Rho-dependent
pathway that stimulates activation of myosin~II, formation of contractile
actin-myosin complexes, and myosin-dependent inhibition of Rac- and
PI3P-dependent responses. Backness signals inhibit frontness signals,
and vice versa (dashed straight lines in Figure~7).
\end{quote}
They go on to explain spontaneous polarization in terms of this kinetic
scheme as follows

\begin{quote}
The more or less symmetrically distributed actin ruffles and PI3P
accumulation seen at early times (e.g., 30 s) after application of
a uniform stimulus presumably mask a fine-textured mosaic of interspersed
backness and frontness signals, some triggering activation of PI3Ks,
Rac, and actin polymerization, others promoting activation of Rho
and myosin. Localized mechanical incompatibility of the two cytoskeletal
responses, combined with the ability of each to damp signals that
promote the other (dashed inhibitor lines in Figure~7), then gradually
drive them to separate into distinct domains of the membrane.
\end{quote}
The goal of this work is to formulate a mathematical model of spontaneous
polarization based on the foregoing mechanism, namely, mutual inhibition
of the frontness and backness pathways.

It turns out that the Bourne model imposes two requirements that are
incompatible with Turing instabilities in a two-component system consisting
of frontness and backness pathways. Specifically, the mutual inhibition
between the frontness and backness pathways must be sufficiently strong
to ensure that the homogeneous steady state is unstable in the presence
of diffusion, and yet weak enough to guarantee its stability in the
absence of diffusion. These two requirements cannot be satisfied simultaneously
in a two-component system, a point that will be discussed in Section~\ref{s:Discussion}.

Thus, we are led to consider the modification of the Bourne model
shown in Figure~\ref{f:ModelScheme}b. It contains two variables,
$U_{2},U_{3}$, representing the frontness and backness pathways,
respectively, whose activation is driven by receptor ligation, and
which inhibit each other in a concentration-dependent manner. It is
assumed furthermore that the mutual inhibition between $U_{2}$ and
$U_{3}$ is so strong that they are mutually incompatible (in a sense
that will be made mathematically precise below). We refer to these
two variables as \emph{activators.} The model differs from the Bourne
picture inasmuch as it assumes the existence of a diffusible \emph{inhibitor},
denoted $U_{1}$, whose synthesis is promoted by both activators,
but which, in turn, inhibits both activators. The inhibitor serves
to stabilize the coexistence of the otherwise incompatible activators:
Transient increases in activator concentrations at localized {}``hotspots''
are efficiently suppressed by the concominant increase in the concentration
of the inhibitor. However, the mobility of the inhibitor imposes constraints
upon its stabilizing effect because it tends to diffuse away from
the {}``hotspot.'' This is not an issue at low chemoattractant concentrations,
for under these conditions, the activator concentrations, and hence,
their mutual incompatibility, is so small that despite the high diffusibility
of the inhibitor, it successfully damps fluctuations of the activator
levels. However, at high chemoattractant concentrations, the activator
concentrations and their mutual incompatibility are so large that
the diffusible inhibitor fails to suppress the two activators sufficiently.
The mutual incompatibility of the two activators now overcomes the
mollifying effect of the inhibitor, and the activators segregate spatially
into separate domains.

The mathematical model in this paper quantifies the above physical
argument. We show that the model yields spontaneous polarization at
sufficiently high active receptor levels. Moreover, if the frontness
or backness pathways are inhibited or activated, they redistribute
spatially in a manner consistent with the experiments described above.
Finally, we show the model displays steady states corresponding to
both chemoattraction and chemorepulsion. Interest in this phenomenon
is motivated by the fact that when neurons are exposed to activators
of the cGMP pathway, chemorepulsion turns into chemorepulsion~\cite{Song1997}.
Based on the analysis of the model, we suggest that such transitions
can be triggered by altering the balance of power in the mutually
inhibitory interactions between the frontness and backness pathways.

The mechanism of spontaneous polarization in this model is distinct
from that in activator-inhibitor or activator-substrate models. Unlike
these models, the spatial segregation of the two activators is driven
entirely by their mutual inhibition --- positive feedback plays no
role. In reality, both positive feedback and mutual inhibition cooperate
to produce symmetry-breaking (Figure~\ref{f:ModelScheme}a). The
contribution of this work is to highlight the distinct role of mutual
inhibition, a feature that emphasized in the experimental literature~\cite{Meili2003,xu03b},
but absent from previous activator-inhibitor and activator-substrate
models.

The paper is organized as follows. In Section~\ref{s:Results}, we
define the model and simulate the experiments described above. In
Section~\ref{s:Discussion}, we elaborate on the physics underlying
the model. Finally, we summarize the conclusions.

\section{\label{s:Results}Results}

\subsection{The model}

We assume that

\begin{enumerate}
\item Both activators promote the synthesis of the inhibitor, which in turn,
degrades by a first-order process. Thus, the net rate of synthesis
of $U_{1}$ is\[
-R_{1}u_{1}+A_{12}u_{2}+A_{13}u_{3}.\]

\item The synthesis of $U_{2}$ is receptor-mediated. It is autocatalytic
at low concentrations and self-limiting at high concentrations, i.e.,\[
R_{2}ru_{2}-A_{22}u_{2}^{2}.\]
where $r$ denotes the receptor activity, and the second term represents
a self-limiting process that prevents synthesis rate from increasing
beyond bounds.
\item The synthesis of $U_{2}$ is inhibited in a concentration-dependent
manner by the common inhibitor, $U_{1}$, and the other activator,
$U_{3}$. Assuming that these interactions follow bimolecular kinetics,
the net rate of synthesis of $U_{2}$ is\[
R_{2}ru_{2}-A_{22}u_{2}^{2}-A_{21}u_{1}u_{2}-A_{23}u_{2}u_{3}.\]

\item The synthesis of $U_{3}$ follows kinetics similar to those of $U_{2}$,
i.e., its net rate of synthesis is\[
R_{3}ru_{3}-A_{33}u_{3}^{2}-A_{31}u_{1}u_{3}-A_{32}u_{2}u_{3}\]

\end{enumerate}
Given these assumptions, we arrive at the equations\begin{align}
\frac{\partial u_{1}}{\partial T} & =D_{1}\frac{\partial u_{1}}{\partial X^{2}}-R_{1}u_{1}+A_{12}u_{2}+A_{13}u_{3}\label{eq:modelEq1}\\
\frac{\partial u_{2}}{\partial T} & =D_{2}\frac{\partial u_{2}}{\partial X^{2}}+\left(R_{2}r-A_{21}u_{1}-A_{23}u_{3}-A_{22}u_{2}\right)u_{2}\label{eq:modelEq2}\\
\frac{\partial u_{3}}{\partial T} & =D_{3}\frac{\partial u_{3}}{\partial X^{2}}+\left(R_{3}r-A_{31}u_{1}-A_{32}u_{2}-A_{33}u_{3}\right)u_{3}\label{eq:modelEq3}\end{align}
where $X$ denotes the spatial coordinate, $T$ denotes time, and
$u_{1},u_{2},u_{3}$ denote the concentrations of $U_{1},U_{2},U_{3}$,
respectively. We assume the Neumann boundary conditions\[
\frac{\partial u_{i}}{\partial X}=0\textnormal{ at }x=0,L\]
since they imply no interaction with the environment, which is consistent
with our desire to study spontaneous (autonomous) polarization.

It is convenient to rescale the equations. If we define\begin{alignat*}{2}
t=a_{11}T, & x=\frac{X}{L}, & d_{i}=\frac{D_{i}/L^{2}}{R_{1}},\\
a_{ij}=\frac{A_{ij}}{R_{i}}, & \rho_{2}=\frac{R_{2}}{R_{1}}, & \rho_{3}=\frac{R_{3}}{R_{1}},\end{alignat*}
we obtain the scaled equations\begin{align}
\frac{\partial u_{1}}{\partial t} & =d_{1}\frac{\partial u_{1}}{\partial x^{2}}-u_{1}+a_{12}u_{2}+a_{13}u_{3}\label{eq:modelDeq1}\\
\frac{\partial u_{2}}{\partial t} & =d_{2}\frac{\partial u_{2}}{\partial x^{2}}+\rho_{2}\left(r-a_{21}u_{1}-a_{22}u_{2}-a_{23}u_{3}\right)u_{2}\label{eq:modelDeq2}\\
\frac{\partial u_{3}}{\partial t} & =d_{3}\frac{\partial u_{3}}{\partial x^{2}}+\rho_{3}\left(r-a_{31}u_{1}-a_{32}u_{2}-a_{33}u_{3}\right)u_{3}\label{eq:modelDeq3}\end{align}
and boundary conditions\[
\frac{\partial u_{i}}{\partial x}=0\textnormal{ at }x=0,1.\]
Our goal is to study the variation of the steady states as a function
of the receptor activity, $r$. We shall show, in particular, that
the model has a homogeneous steady state at which $u_{2}$ and $u_{3}$
coexist, but it becomes Turing unstable at a sufficiently large $r$.
The non-homogeneous steady state emerging from the Turing bifurcation
is such that $u_{2}$ and $u_{3}$ are spatially segregated.

\subsection{The homogeneous steady states}

\begin{figure}
\begin{center}\subfigure[]{\includegraphics[%
  width=2.7in,
  height=2.7in]{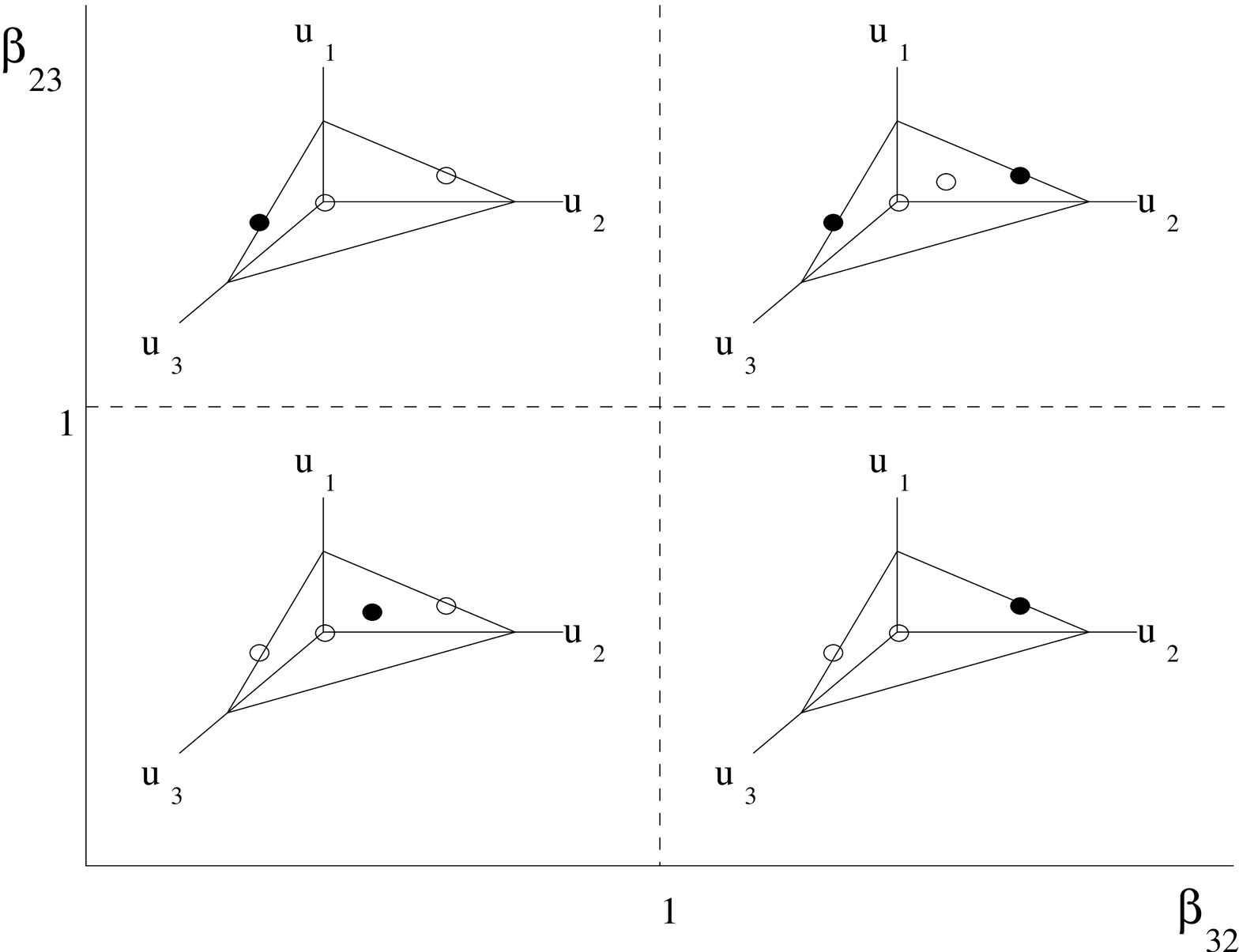}}\hspace{1in}\subfigure[]{\includegraphics[%
  width=2.7in,
  height=2.7in]{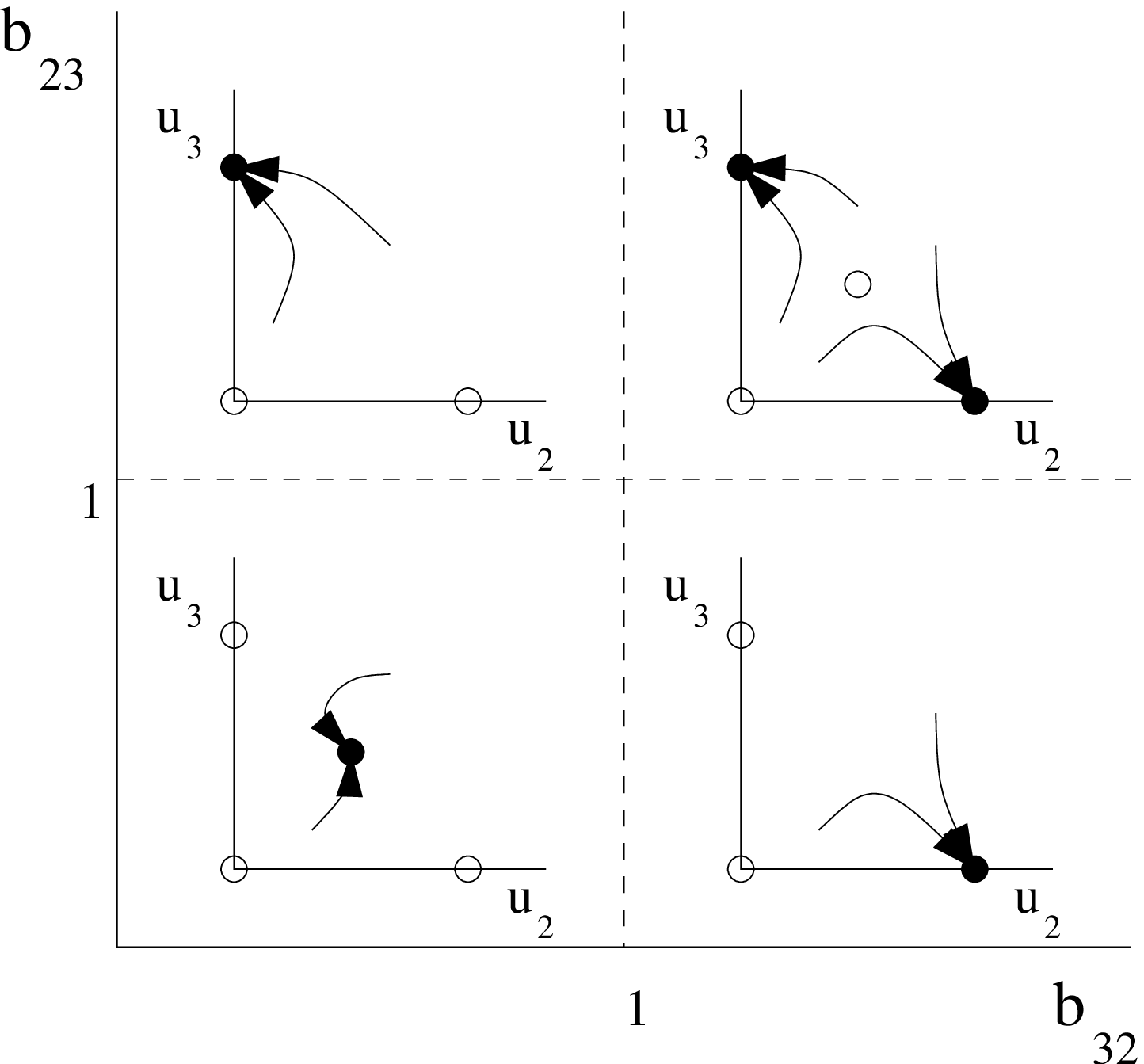}}\end{center}

\caption{\label{f:BifurcationDiagram}The bifurcation diagram for the homogeneous
steady states (a) In the presence of the inhibitor and (b)~in the
absence of the inhibitor. The stable and unstable steady states are
represented by full and open circles, respectively.}
\end{figure}

A Turing instability occurs when a non-homogeneous steady state bifurcates
from a homogeneous steady state that is stable in the absence of diffusion.
It is therefore useful to characterize the conditions for stability
of the homogeneous steady states in the absence of diffusion.

The model has three types of homogeneous steady states: the trivial
steady state, $u_{1}=u_{2}=u_{3}=0$, denoted $E_{000}$; the semitrivial
steady states, $u_{1},u_{2}>0,u_{3}=0$, and $u_{1},u_{3}>,u_{2}=0$,
denoted $E_{110}$ and $E_{101}$, respectively; and the nontrivial
steady state, $u_{1},u_{2},u_{3}>0$, denoted $E_{111}$. The key
results, depicted graphically in Figure~\ref{f:BifurcationDiagram}a,
are as follows (see Appendix~\ref{a:HSSstability} for details):

\begin{enumerate}
\item The trivial steady state, $E_{000}$, exists for all $r>0$, and is
always unstable.
\item The semi-trivial steady state, $E_{110}$, which lies on the $u_{1}u_{2}$-plane,
exists for all $r>0$. It is stable if and only if\[
\beta_{32}\equiv\frac{\alpha_{32}}{\alpha_{22}}>1.\]
Here, $\alpha_{22}\equiv a_{22}+a_{21}a_{12}$ is a measure of the
extent to which $U_{2}$ inhibits itself both directly ($a_{22}$)
and indirectly through production of $U_{1}$ ($a_{21}a_{12}$). Likewise,
$\alpha_{32}\equiv a_{32}+a_{31}a_{12}$ is a measure of the extent
to which $U_{2}$ inhibits $U_{3}$ directly ($a_{32}$) and indirectly
through production of $U_{1}$ ($a_{31}a_{12})$. The foregoing stability
condition then says that $E_{110}$ is stable if and only if $U_{2}$
inhibits $U_{3}$ more than it inhibits itself.
\item The semi-trivial steady state, $E_{101}$, which lies on the $u_{1}u_{3}$-plane,
exists for all $r>0$. It is stable if and only if \[
\beta_{23}\equiv\frac{\alpha_{23}}{\alpha_{33}}>1,\]
where $\alpha_{23}\equiv a_{23}+a_{21}a_{13},\alpha_{33}\equiv a_{33}+a_{31}a_{13}$
are measures of the extent to which $U_{3}$ inhibits $U_{2}$ and
itself, respectively, and $\beta_{23}$ is measure of the cross-inhibition
relative to the self-inhibition.
\item The coexistence steady state, $E_{111}$, exists if and only if\[
\beta_{23},\beta_{32}<1\;\textnormal{ or }\;\beta_{23},\beta_{32}>1.\]
It is stable only if the mutual inhibition of $U_{2}$ and $U_{3}$
is sufficiently weak, i.e., $\beta_{23}\beta_{32}<1$.
\end{enumerate}
Figure~\ref{f:BifurcationDiagram}a is reminiscent of the bifurcation
diagram for the Lotka-Volterra model~\cite{murray}. This is not
surprising because $U_{2}$ and $U_{3}$ obey Lotka-Volterra dynamics
in the absence of the inhibitor. Indeed, the equations obtained by
letting $u_{1}=0$ in equations (\ref{eq:modelDeq2}--\ref{eq:modelDeq3})
are identical to the Lotka-Volterra model for two competing species.
It follows that the dynamics of $U_{2}$ and $U_{3}$ in the absence
of the inhibitor are described by a bifurcation diagram very similar
to Figure~\ref{f:BifurcationDiagram}a, the only difference being
that $\beta_{23}$ and $\beta_{32}$ must now be replaced by $b_{23}\equiv a_{23}/a_{22},\; b_{32}\equiv a_{32}/a_{33}$
(see Figure~\ref{f:BifurcationDiagram}b). We shall appeal to this
fact below.

\subsection{Turing instability of homogeneous steady states}

Since $E_{000}$ is always unstable, it can never undergo a Turing
bifurcation. However, the semi-trivial and non-trivial steady states
are stable for all sufficiently small $r>0$. The question then arises
whether these steady states can undergo a Turing bifurcation. It is
shown in Appendix~\ref{a:HSSturingStability} that

\begin{enumerate}
\item The semi-trivial steady states cannot undergo a Turing bifurcation.
\item The nontrivial steady state, $E_{111}$, can undergo a Turing bifurcation,
but this is so only if \begin{equation}
b_{23}b_{32}>1,\label{eq:NCforTuringInstability}\end{equation}
i.e., the mutual inhibition between $U_{2}$ and $U_{3}$ must be
sufficiently strong --- so strong, in particular, that they cannot
coexist in the absence of the inhibitor (see Figure~\ref{f:BifurcationDiagram}b).
\end{enumerate}
Both conclusions are a consequence of the following fact which will
be discussed in Section~\ref{s:Discussion}: In this model, the only
destabilizing mechanism driving the Turing instability is mutual inhibition
of $U_{2}$ and $U_{3}$, which requires the existence of \emph{both}
activators. Thus, the semi-trivial steady states fail to undergo a
Turing bifurcation because one of the two activators is absent at
such steady states. The non-trivial steady state, which is characterized
by positive concentrations of $U_{2}$ and $U_{3}$, allows for a
Turing instability, but only if their mutual inhibition is sufficiently
strong.

Although the mutual inhibition must be sufficiently strong to ensure
that $E_{111}$ undergoes a Turing instability, intuition suggests
that it cannot be too strong, lest the two activators become incompatible
even in the presence of the inhibitor. This is indeed the case. To
see this, observe that (\ref{eq:NCforTuringInstability}) can be satisfied
in three different ways (see Figure~\ref{f:BifurcationDiagram}b):

\begin{enumerate}
\item $b_{23},b_{32}>1$, i.e., the mutual inhibition between $U_{2}$ and
$U_{3}$ is so strong that they display bistable dynamics in the absence
of the inhibitor.
\item $b_{23}<1,b_{32}>1$, i.e., in the absence of the inhibitor, $U_{2}$
inhibits $U_{3}$ much more than $U_{3}$ inhibits $U_{2}$, so that
$U_{2}$ ultimately prevails over $U_{3}$.
\item $b_{23}>1,b_{32}<1$, i.e., in the absence of the inhibitor, $U_{3}$
inhibits $U_{2}$ much more than $U_{2}$ inhibits $U_{3}$, so that
$U_{3}$ ultimately prevails over $U_{2}$.
\end{enumerate}
It turns out that in the first case, when both $U_{2}$ and $U_{3}$
inhibit each other strongly, the necessary condition (\ref{eq:NCforTuringInstability})
is satisfied. Yet, the Turing instability cannot be realized because
a stable coexistence steady state fails to exist even in the presence
of the inhibitor. This follows from the fact that $E_{111}$ exists
and is stable only if the mutual inhibition is sufficiently weak in
the presence of the inhibitor, i.e.,\[
\beta_{23}=\frac{a_{23}+a_{21}a_{13}}{a_{33}+a_{31}a_{13}}<1,\;\beta_{32}=\frac{a_{32}+a_{31}a_{12}}{a_{22}+a_{21}a_{12}}<1\]
which can be recast in the form\begin{equation}
\frac{a_{22}}{a_{12}}\left(b_{32}-1\right)<a_{21}-a_{31}<\frac{a_{33}}{a_{13}}\left(1-b_{23}\right).\label{eq:ExistenceOfE111b}\end{equation}
Clearly, (\ref{eq:ExistenceOfE111b}) cannot be satisfied if $b_{23},b_{32}>1$.
Under these conditions, the mutual inhibition of $U_{2}$ and $U_{3}$
is so strong that they cannot coexist stably even in the presence
of the inhibitor.

We conclude that $E_{111}$ exists and bifurcates via a Turing instability
only if the interaction between $U_{2}$ and $U_{3}$ in the absence
of the inhibitor is such that only one of them prevails ultimately.
Furthermore

\begin{enumerate}
\item If $U_{2}$ prevails over $U_{3}$ in the absence of the inhibitor
($b_{23}<1,b_{32}>1$), then a Turing instability obtains only if
$b_{23}b_{32}>1$ and \begin{equation}
0<\frac{a_{22}}{a_{12}}\left(b_{32}-1\right)<a_{21}-a_{31}<\frac{a_{33}}{a_{13}}\left(1-b_{23}\right).\label{eq:NCu2Wins}\end{equation}

\item Conversely, if $U_{3}$ prevails over $U_{2}$ in the absence of the
inhibitor ($b_{23}>1,b_{32}<1$), then a Turing instability obtains
only if $b_{23}b_{32}>1$ and\begin{equation}
\frac{a_{22}}{a_{12}}\left(b_{32}-1\right)<a_{21}-a_{31}<\frac{a_{33}}{a_{13}}\left(1-b_{23}\right)<0.\label{eq:NCu3Wins}\end{equation}

\end{enumerate}
These conditions have a simple physical interpretation. Consider,
for instance, the condition~(\ref{eq:NCu2Wins}). It says that if
$U_{2}$ prevails over $U_{3}$ in the absence of the inhibitor, a
Turing instability obtains only if (a) $a_{21}-a_{31}>0$, i.e., $U_{1}$
inhibits $U_{2}$ more than it inhibits $U_{3}$. The stronger inhibition
of $U_{2}$ by $U_{1}$ is necessary in order to compensate for its
intrinsic superiority over $U_{3}$. (b) The magnitude of this difference
must be neither too small nor too large to prevent under- or over-compensation
that would preclude the coexistence of $U_{2}$ and $U_{3}$ even
in the presence of the inhibitor.

It is shown in Appendix~\ref{a:HSSturingStability} that the conditions
(\ref{eq:NCu2Wins}) or (\ref{eq:NCu3Wins}) are not only necessary
but almost sufficient for $E_{111}$ to undergo a Turing instability.
It suffices to impose the additional condition that $d_{1}$ be sufficiently
larger than $d_{2}$ and $d_{3}$.

\subsection{Simulation of experiments}

\begin{table}

\caption{\label{t:ParameterValues}Parameter values for the simulations. The
diffusion coefficient of the inhibitor, $D_{1}$, is assumed to much
larger than the diffusion coefficients, $D_{2},D_{3}$, of the frontness
and backness pathways. The rate constants, $R_{i},A_{ij}$, were chosen
in order to satisfy the conditions~(\ref{eq:NCu2Wins}) or (\ref{eq:NCu3Wins}).
These two cases are shown in the columns labeled column labeled {}``$U_{2}$
wins'' and {}``$U_{3}$ wins'', respectively.}

\begin{center}\begin{tabular}{|c|c|c|c|c|c|}
\hline
Parameter&
$U_{2}$ wins&
$U_{3}$ wins&
Parameter&
$U_{2}$ wins&
$U_{3}$ wins\tabularnewline
\hline
\hline
$D_{1}$&
3&
3&
$A_{13}$&
1/3&
4\tabularnewline
\hline
$D_{2}$&
0.001&
0.001&
$A_{21}$&
3&
1\tabularnewline
\hline
$D_{3}$&
0.005&
0.005&
$A_{22}$&
1&
1\tabularnewline
\hline
$R_{1}$&
1&
1&
$A_{23}$&
1&
2\tabularnewline
\hline
$R_{2}$&
2&
1&
$A_{31}$&
1&
3\tabularnewline
\hline
$R_{3}$&
1&
2&
$A_{32}$&
2&
1\tabularnewline
\hline
$A_{12}$&
4&
1/2&
$A_{33}$&
1&
1\tabularnewline
\hline
\end{tabular}\end{center}
\end{table}

\begin{figure}
\begin{center}\subfigure[]{\includegraphics[%
  width=3in,
  height=2in]{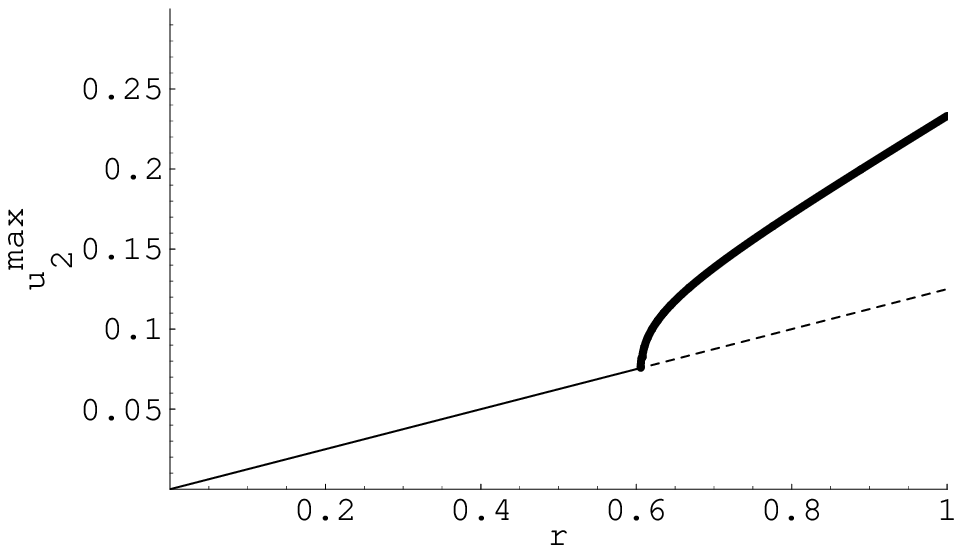}}\hspace{0.1in}\subfigure[]{\includegraphics[%
  width=3in,
  height=2in]{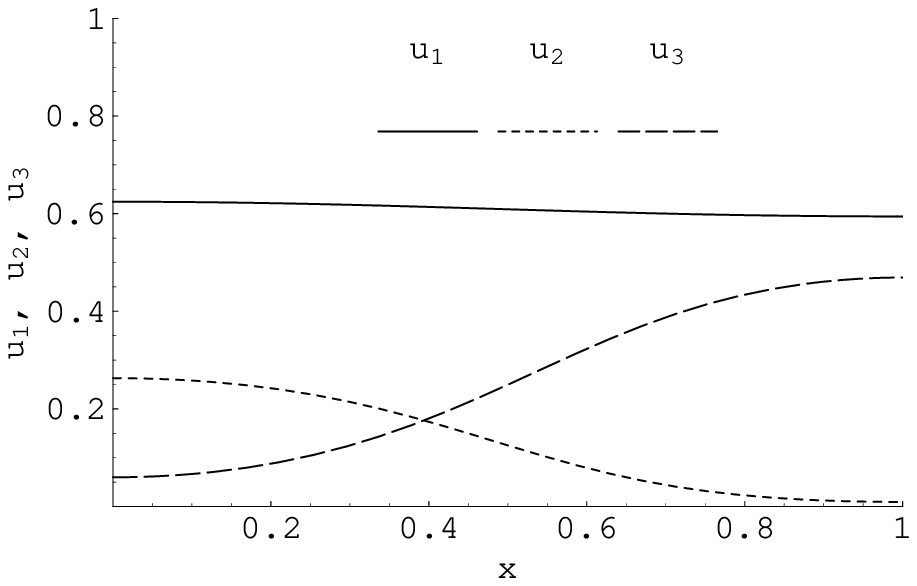}}\end{center}

\caption{\label{f:NHSS_u2Wins}Symmetry-breaking of the non-trivial steady
state when the parameters are chosen such that $U_{2}$ wins in the
absence of the inhibitor. (a)~The bifurcation diagram depicting the
variation of $u_{2}^{\textnormal{max}}\equiv\max_{0\le x\le1}u_{2}(x)$
as a function of the active receptor concentration, $r$. The thin
line represents the homogeneous non-trivial steady state, $E_{111}$;
the dashed portion of this line indicates that $E_{111}$ is Turing
unstable. The thick line represents the non-homogeneous steady state
emerging from the Turing bifurcation at $r\approx0.6$. (b) The profiles
of $u_{1}$, $u_{2}$, and $u_{3}$ at the non-homogeneous steady
state corresponding to $r=1$. The steady state was attained by perturbing
the homogeneous steady state at $r=1$, i.e., by integrating equations
(\ref{eq:modelDeq1}--\ref{eq:modelDeq3}) with initial conditions,
$u_{1}(0,x)=\tilde{u}_{1}$, $u_{i}(0,x)=\tilde{u}_{i}[1+0.01\cos(\pi x)],i=2,3$,
where $\tilde{u_{i}}=1,2,3$ denotes the non-trivial homogeneous steady
state at $r=1$.}
\end{figure}

To simulate the data shown in Figure~\ref{f:SpatialSegregation}b,
wherein the frontness and backness components segregate spontaneously
in response to a uniform chemoattractant profile, we chose parameter
values satisfying (\ref{eq:NCu2Wins}), shown in the column labeled
{}``$U_{2}$ wins'' of Table~\ref{t:ParameterValues}. Linear stability
analysis shows that given these parameter values, the coexistence
steady state, $E_{111}$, undergoes a Turing instability at $r\approx0.6$
and wavenumber $k\approx1$ (Figure~\ref{f:detCk}a). Computations
with the continuation software package CONTENT~\cite{content} confirm
the existence of this instability (Figure~\ref{f:NHSS_u2Wins}a).
The homogeneous steady state is stable for $0<r\lesssim0.6$, and
Turing unstable therafter. The concentration profiles of the non-homogeneous
steady state created at the Turing bifurcation point are consistent
with the data. Figure~\ref{f:NHSS_u2Wins}b shows that the non-homogeneous
steady state at $r=1$ is such that $U_{2}$ and $U_{1}$ are in phase,
and $U_{2}$ and $U_{3}$ are out of phase. The latter is consistent
with the data in Figure~\ref{f:SpatialSegregation}b. We refer to
this steady state as the \emph{chemoattraction} steady state for the
following reason. To a first degree of approximation, the steady state
obtained in the presence of a receptor gradient, $r(x)=1+\epsilon\cos(\pi x),0<\epsilon\ll1$,
is identical to the steady state shown in Figure~\ref{f:NHSS_u2Wins}b.
Thus, the profile of the frontness pathways is in phase with the distribution
of the active receptors, which is characteristic of chemoattraction.

\begin{figure}
\begin{center}\subfigure[]{\includegraphics[%
  width=2in,
  height=1.5in]{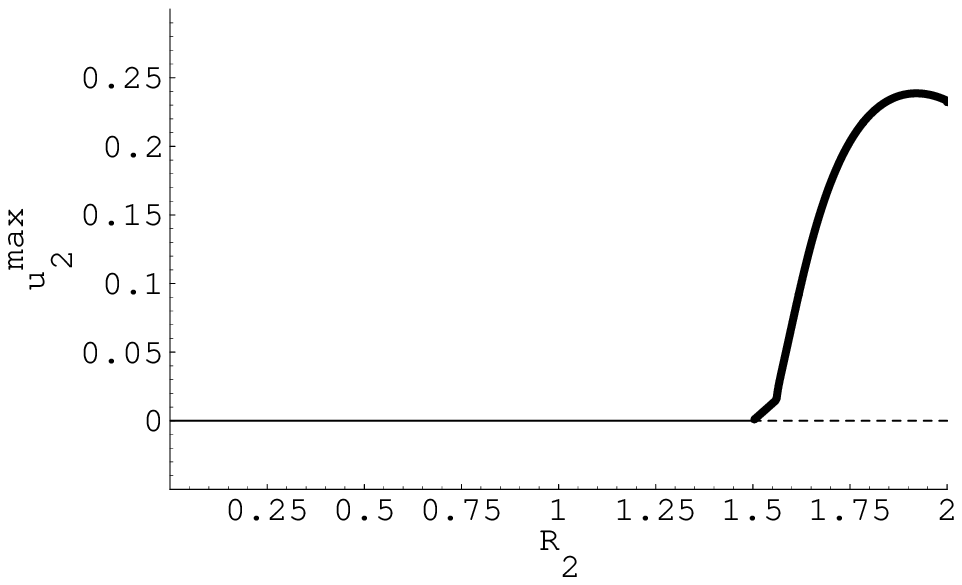}}\hspace{0.1in}\subfigure[]{\includegraphics[%
  width=2in,
  height=1.5in]{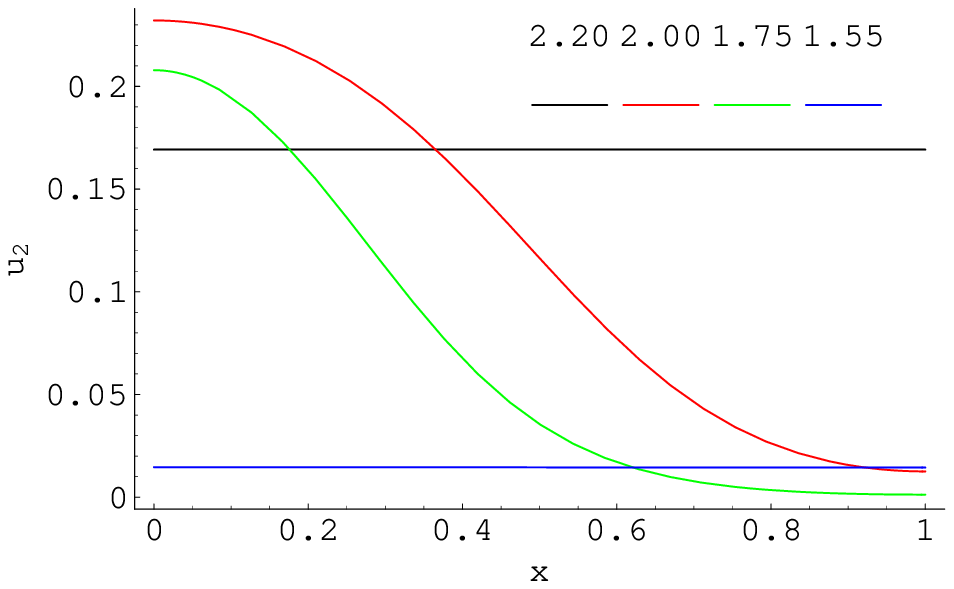}}\hspace{0.1in}\subfigure[]{\includegraphics[%
  width=2in,
  height=1.5in]{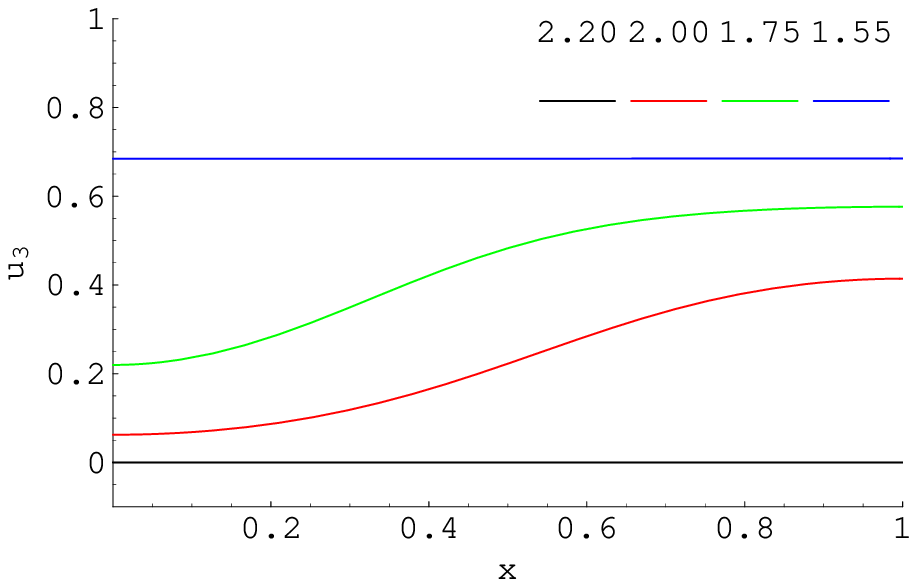}}\end{center}

\caption{\label{f:NHSS_r2Variation}Variation of the chemoattraction steady
state with respect to $R_{2}$ at $r=1$. (a)~The bifurcation diagram
showing the variation of $u_{2}^{\textnormal{max}}$ as a function
of $R_{2}$. The thick line represents the non-homogeneous steady
state. The thin line represents the homogeneous semi-trivial steady
state, $E_{101}$. As $R_{2}$ decreases, so does $u_{2}^{\textnormal{max}}$
until it intersects the branch, $u_{2}^{\textnormal{max}}=0$, which
corresponds the homogeneous semi-trivial steady state, $E_{101}$.
(b)~As $R_{2}$ decreases, the concentration profile of $U_{2}$
decreases in magnitude and occupies a progressively smaller proportion
of the cell until $u_{2}$ is identically zero. (c)~As $R_{2}$ decreases,
the concentration profile of $U_{3}$ increases in magnitude and advances
into the pre-existing front until $u_{3}$ acquires the uniform profile
of $E_{101}$.}
\end{figure}

Figures~\ref{f:FrontnessInhibitionActivation} and \ref{f:Chemorepulsion}
show the redistribution of the frontness and backness pathways in
response to inhibition and activation of the frontness pathways. To
simulate these experiments, we computed the variation of the chemoattraction
steady state at $r=1$ with respect to $R_{2}$. Figure~\ref{f:NHSS_r2Variation}a
shows that as $R_{2}$ decreases, so does $u_{2}^{\textnormal{max}}$
until it becomes zero, i.e., the chemoattraction steady state ceases
to exist, and is replaced by the homogeneous steady state, $E_{101}$,
in which $u_{2}=0$ throughout the spatial domain. The variation of
the concentration profile for the frontness pathway is consistent
with the data shown in Figure~\ref{f:FrontnessInhibitionActivation}.
Figures~\ref{f:NHSS_r2Variation}b shows that as $R_{2}$ decreases,
so does the gradient of the frontness molecules until it vanishes
at $R_{2}\approx1.5$, which is in qualitative agreement with the
data shown in Figure~\ref{f:FrontnessInhibitionActivation}a. Conversely,
at sufficiently large values of $R_{2}$, the non-homogeneous steady
state is replaced by the homogeneous steady state, $E_{110}$, characterized
by high levels of the frontness molecules uniformly distributed throughout
the spatial domain (see Figure~\ref{f:FrontnessInhibitionActivation}b).
The variation of the concentration profile for the backness pathway
is also consistent with the data. Figures~\ref{f:NHSS_r2Variation}c
shows that as $R_{2}$ decreases, the backness pathway progressively
advances into the pre-existing front until it occupies not only the
back but also the front of the cell. This is consistent with the data
shown in Figure~\ref{f:Chemorepulsion}a.

Simulations in which $r$ is held at 1.0 and $R_{3}$ is progressively
changed give analogous results (simulations not shown). As $R_{3}$
increases, the frontness pathways occupy a progressively smaller portion
of the cell until there is no frontness throughout the cell. The simulations
did not yield multiple peaks corresponding to the multiple pseudopods
shown in Figure~\ref{f:BacknessInhibitionActivation}b. Numerical
calculations of the critical wavenumber, $k_{0}$, show that it increases
as $R_{3}$ decreases, which is the correct trend. However, the steady
state ceases to exist before $k_{0}$ can exceed 2.

\begin{figure}
\begin{center}\subfigure[]{\includegraphics[%
  width=3in,
  height=2in]{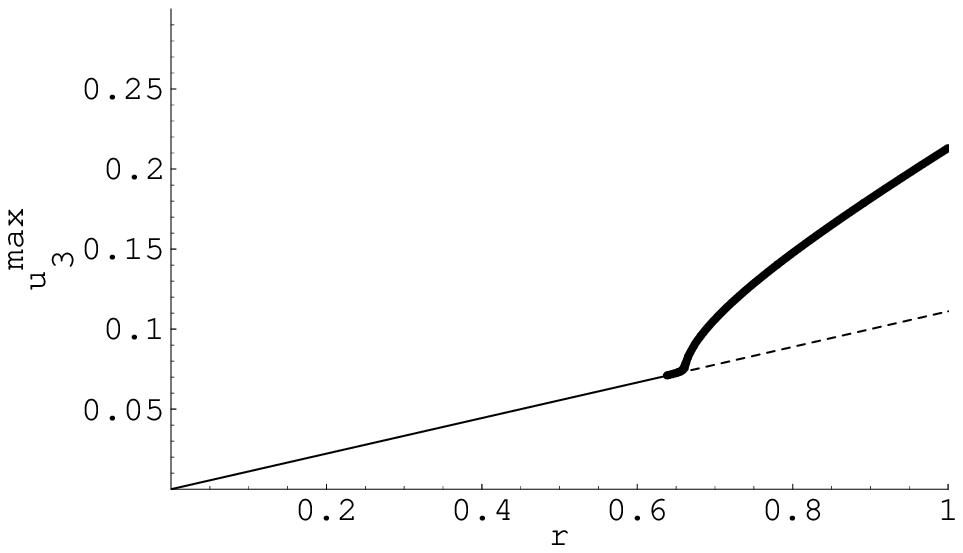}}\hspace{0.1in}\subfigure[]{\includegraphics[%
  width=3in,
  height=2in]{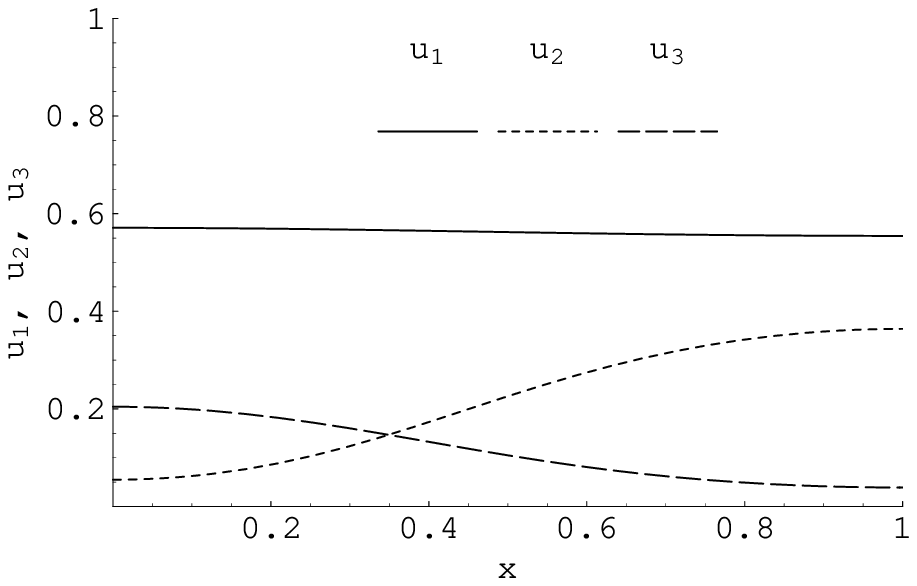}}\end{center}

\caption{\label{f:NHSS_u3Wins}Symmetry-breaking of the non-trivial steady
state when the parameters are chosen such that $U_{3}$ wins in the
absence of the inhibitor. (a)~The bifurcation diagram depicting the
variation of $u_{3}^{\textnormal{max}}\equiv\max_{0\le x\le1}u_{3}(x)$
as a function of the active receptor concentration, $r$. The thin
line represents the homogeneous non-trivial steady state, $E_{111}$;
the dashed portion of this line indicates that $E_{111}$ is Turing
unstable. The thick line represents the non-homogeneous steady state
emerging from the Turing bifurcation at $r\approx0.65$. (b) The profiles
of $u_{1}$, $u_{2}$, and $u_{3}$ at the non-homogeneous steady
state attained by perturbing the homogeneous steady state at $r=1$.}
\end{figure}

To investigate the existence of steady states corresponding to chemorepulsion,
we chose parameter values satisfying (\ref{eq:NCu3Wins}) so that
$U_{3}$ prevails over $U_{2}$ in the absence of the inhibitor (column
labeled {}``$U_{3}$ wins'' of Table~\ref{t:ParameterValues}).
Here, the coexistence steady state, $E_{111}$, undergoes a Turing
bifurcation at $r\approx0.65$ and $k=1$ (Figures~\ref{f:detCk}b).
The non-homogeneous steady state that emerges from this bifurcation
is such that $U_{2}$ and $U_{3}$ are once again out of phase,%
\footnote{It is shown in~Appendix~\ref{a:HSSturingStability} that regardless
of the parameter values, $U_{2}$ and $U_{3}$ will always be out-of-phase
in the non-homogeneous steady state bifurcating from $E_{111}$.%
} but it is $U_{3}$, instead of $U_{2}$, that is in phase with $U_{1}$
(Figure~\ref{f:NHSS_u3Wins}b). This non-homogeneous steady state
represents the \emph{chemorepulsion} steady state, since it approximates
the steady state that would be obtained in the presence of a receptor
gradient, $r(x)=1+\epsilon\cos(\pi x),0<\epsilon\ll1$, and the profile
of the frontness pathways are, in this case, out of phase with the
active receptor distribution. Given the symmetry of the equations,
it is not surprising that the model admits a chemorepulsion steady
state. For, if choose the parameters corresponding to the column {}``$U_{2}$
wins'' in Table~\ref{t:ParameterValues}, but interchange $u_{2}$
and $u_{3}$, the chemoattraction steady state in Figure~\ref{f:NHSS_u2Wins}b
turns into a chemorepulsion steady state. However, the analysis of
the model suggests a mechanism for the transition. Specifically, the
conditions (\ref{eq:NCu2Wins}--\ref{eq:NCu3Wins}) imply that transitions
from chemorepulsion to chemattraction, such as those observed in neurons~\cite{Song1997},
can be triggered by pharmacological agents that selectively inhibit
the backness pathway so that the frontness pathway becomes intrinsically
superior in the absence of the inhibitor.

\section{\label{s:Discussion}Discussion}

We have shown that a model based on mutual inhibition of the frontness
and backness pathways can yield spontaneous polarization involving
spatial segregation of the two pathways. The frontness and backness
pathways are mutually incompatible in the absence of the inhibitor
(eq.~\ref{eq:NCforTuringInstability}), but coexist in the presence
of the inhibitor because it suppresses the growth, and hence, the
mutual incompatibility, of the frontness and backness pathways. Since
the inhibitor is diffusible, it can suppress the nominal incompatibility
manifested at low receptor levels. However, it fails to achieve this
goal at high receptor levels, resulting in spatial segregation of
the frontness and backness pathways (see Figures~\ref{f:SpatialSegregation}
and~\ref{f:NHSS_u2Wins}).

Since the model takes due account of the backness pathways, we could
explore the effect of frontness inhibition or activation on the backness
pathways, and the effect of backness modification on the frontness
and backness pathways. The simulations in Section~\ref{s:Results}
yielded results that are in qualitative agreement with the data shown
in Figures~\ref{f:SpatialSegregation}--\ref{f:BacknessInhibitionActivation}.
These results are beyond the scope of activator-inhibitor or activator-substrate
models of gradient sensing since they do not account for the backness
pathways.

The particular manner in which the two pathways segregate depends
on their behavior in the absence of the tempering influence of the
inhibitor. If the frontness pathways dominate over the backness pathways,
their spatial segregation is consistent with chemoattraction. Conversely,
if the backness pathways dominate, the emergent spatial pattern corresponds
to chemorepulsion. The model therefore suggests that transitions between
chemoattraction and chemorepulsion can be provoked by agents that
shift the balance of power between the frontness and backness pathways.

The model proposed here is the simplest possible model of spontaneous
polarization driven by mutual inhibition. By this, we mean the following

\begin{enumerate}
\item A model containing a smaller number of variables will not yield spontaneous
polarization.
\item The model is simpler than any other model in which the symmetry-breaking
is driven entirely by mutual inhibition.
\end{enumerate}
In what follows, we justify these conclusions, since they are required
to explain certain points raised above.

To see that a model containing 2 variables will not yield spontaneous
polarization driven by mutual inhibition, let $u=[u_{1},u_{2}]^{t}$
be a homogeneous steady state of a two-component reaction-diffusion
system\[
\frac{\partial u_{i}}{\partial t}=D_{i}\frac{\partial u_{i}}{\partial x^{2}}+f_{i}(u_{1},u_{2}),\; i=1,2\]
 and let $J(u)$ denote the Jacobian at $u$. Then, $u$ can undergo
a Turing instability only if~\cite{segel72}\begin{equation}
J_{11}+J_{22}<0,\; J_{11}J_{22}-J_{12}J_{21}>0,\; D_{1}J_{22}+D_{2}J_{11}>0.\label{eq:TuringConditions}\end{equation}
The first and third conditions imply that $J_{11}J_{22}<0$, i.e.,
one of the components, say $U_{1}$, must inhibit its own synthesis
($J_{11}<0$), and the other, say $U_{2}$, must activate its own
synthesis ($J_{22}>0$). Now, if the two components inhibit each other,
then $J_{12},J_{21}<0$, so that $J_{12}J_{21}>0$ and $J_{11}J_{22}-J_{12}J_{21}<0$,
which violates the second condition for Turing stability. Thus, a
two-component model in which each component inhibits the other component
cannot yield spontaneous polarization. Indeed, such a model can yield
a Turing instability only if $J_{12}J_{21}<0$, i.e., the interaction
between $U_{1}$ and $U_{2}$ is asymmetric. If $U_{1}$ inhibits
$U_{2}$, then $U_{2}$ must activate $U_{1}$; this corresponds to
the \emph{activator-inhibitor} model; conversely, if $U_{1}$ activates
$U_{2}$, then $U_{2}$ must inhibit $U_{1}$, which corresponds to
the \emph{activator-substrate} model. No other cross-interactions
are admissible. It follows that if spontaneous polarization is to
be driven by mutual inhibition, there must be at least three components.
Thus, we were led to modify the Bourne model by postulating the existence
of an additional component, namely, a diffusible inhibitor.

To show that the model is simpler than any other model in which symmetry-breaking
is driven entirely by mutual inhibition, we appeal to the fact that
in a general reaction-diffusion system, a homogeneous steady state
cannot become Turing instable unless it contains an \emph{unstable
subsytem}~\cite{Satnoianu2000}.%
\footnote{A \emph{subsystem} refers to any proper subset of the model variables.
For example, the two-component model discussed above has two subsystems,
$\{ U_{1},U_{2}\}$. A subsystem is said to be \emph{unstable} if
it becomes unstable as soon as the remaining components of the system
are somehow rendered constant. In other words, without the stabilizing
effects of the other variables, the subsystem diverges from its values
at the homogeneous steady state. In the 2-component model, the activator
is an unstable subsystem: If the inhibitor or substrate level is somehow
held constant, the activator diverges from the homogeneous steady
state due to positive feedback ($J_{22}>0$). As shown above, the
existence of such an unstable subsystem is necessary for the full
system to undergo a Turing instability.%
} The three-component model considered here has three 1-component subsystems,
$\{ U_{1},U_{2},U_{3}\}$, and three 2-component subsystems, $\{ U_{1}U_{2},U_{2}U_{3},U_{1}U_{3}\}$.
It turns out that in the neighborhood of the non-trivial steady state,
$E_{111}$, all three 1-component subsystems are stable, i.e. $J_{11},J_{22},J_{33}<0$
(see~\ref{eq:JacobianE111}). In other words, the self-interactions
of the individual components are stabilizing. Notably, $U_{2}$ and
$U_{3}$ are stable subsystems even though their synthesis was assumed
to be autocatalytic. This is because the kinetics of their synthesis
are such that the feedback is positive only when $u_{2}$ and $u_{3}$
are small. However, in the neighborhood of $E_{111}$, $u_{2}$ and
$u_{3}$ are so large that the self-limiting effect dominates, and
the feedback is, in fact, negative. Now, since the 1-component subsystems
are stable, $E_{111}$ can be Turing unstable only if at least one
of the 2-component subsystems is unstable, i.e., the cross-interaction
of at least one 2-component system is destabilizing. Such destabilizing
cross-interactions can occur only if the interaction between the 2
components is mutually inhibitory or synergistic, i.e.,\[
J_{12}J_{21}>0\textnormal{ or }J_{13}J_{31}>0\textnormal{ or }J_{23}J_{32}>0.\]
 The first two conditions cannot be satisfied since the interaction
between $U_{1}$ and each of the two activators, $U_{2}$ and $U_{3}$
is neither mutually inhibitory nor mutually synergistic --- $U_{1}$
inhibits $U_{2}$ and $U_{3}$, but $U_{2}$ and $U_{3}$ activate
$U_{1}$. Thus, $\{ U_{1},U_{2}\}$ and $\{ U_{1},U_{3}\}$ are stable
subsystems, and $\{ U_{2},U_{3}\}$ is the only unstable subsystem
in the model. Consequently, the spatial segregation of the frontness
and backness pathways is driven \emph{entirely} by mutual inhibition
between $U_{2}$ and $U_{3}$ --- positive feedback plays no role
since $J_{11},J_{22},J_{33}<0$.

It should be noted that reciprocal distribution of the frontness and
backness pathways (specifically, PI3K and PTEN) is a central feature
of the local-excitation-global-inhibition model proposed by Iglesias
and coworkers~\cite{Ma2004}. However, the mechanism for spatial
segregation of PI3K and PTEN is such that polarization occurs only
in the presence of a chemoattractant gradient. Indeed, according to
this model

\begin{enumerate}
\item Exposure to chemoattractant generates binding sites for PI3K and PTEN
at the membrane.
\item The rate of generation of binding sites for PI3K and PTEN is proportional
to $r^{2}$ and $r$, respectively, where $r$ denotes the active
receptor concentration.
\item The concentration of PI3P is determined by the relative concentrations
of membrane-bound PI3K and PTEN --- the higher the PI3K:PTEN ratio,
the larger the concentration of PI3P.
\end{enumerate}
It follows from (2) that when a cell is exposed to a gradient, the
leading edge develops a high PI3K:PTEN ratio, while the trailing edge
has a high PTEN:PI3K ratio. Then, (3) implies that the PI3P concentration
at the leading edge is higher than that at the trailing edge. Importantly,
the polarized distribution of PI3P arises only in the presence of
a gradient. For, if $r$ is constant, the PI3K:PTEN ratio, and hence,
the concentration of PI3P, is constant all over the cell membrane.
Thus, the local-excitation-global-inhibition model cannot capture
the spontaneous polarization observed in the absence of gradients.

Although our model captures spontaneous polarization driven by mutual
inhibition of frontness and backness pathways, it is missing two important
features of the experimental data. First, it is known that both negative
\emph{and} positive feedback operate in chemoattractant-mediated polarization
(see Figure~\ref{f:ModelScheme}a and refs.~\cite{Meili2003,xu03b}).
Although we assumed the existence of positive feedback in the model,
it played no role in the spontaneous polarization since $J_{22},J_{33}<0$
in the neighborhood of $E_{111}$. Second, the model does not account
for adapation, the process that enables the cell to ultimately return
to the homogeneous steady state even in the presence of chemoattractant~\cite{parent97}.
A modified model taking due account of positive feedback and long-term
adaptation is currently under investigation.

Two of the model variables ($U_{2},U_{3}$) correspond to the frontness
and backness pathways of the Bourne scheme. However, the model hypothesizes
the existence of an additional variable, namely, a diffusible inhibitor
required to suppress the mutual incompatibility of the frontness and
backness pathways. It is, by no means, necessary that this diffusible
component be an inhibitor. It could just as well be a diffusible \emph{substrate}
that is required for the synthesis of the two activators, but is consumed
in process of activator synthesis. This would reverse the signs of
$J_{12},J_{13},J_{21},J_{31}$, but the key inequalities in the above
arguments, $J_{13}J_{31},J_{12}J_{21}<0$, would be preserved. A diffusible
inhibitor was assumed only because there is some evidence of their
existence.

\begin{enumerate}
\item Luo and coworkers have shown that cytosolic inositol phosphates, whose
levels surge immediately after chemoattractant stimulation, inhibit
the frontness pathways in both \emph{Dictyostelium}~\cite{luo03}
and neutrophil-like HL-60 cells~\cite{Luo}.
\item Nimnual \emph{et al} have shown that Rac-mediated production of reactive
oxygen species downregulates the backness component, Rho~\cite{Nimnual2003}.
\end{enumerate}
It remains to be seen if these diffusible inhibitors play a critical
role in the maintenance of polarity, as required by the diffusible
inhibitor of the proposed model.

\section{Conclusions}

We have formulated a model that provides a mathematical realization
of the Bourne scheme for spontaneous polarization by mutual inhibition
of frontness and backness pathways~\cite{xu03b}. The model predicts
several experimentally observed features that are outside the scope
of prevailing models of gradient sensing.

\begin{enumerate}
\item Mutual inhibition of the frontness and backness pathways plays a critical
role in generating the spontaneous polarization.
\item If the frontness pathway is suppressed, the backness pathways occupy
a progressively larger proportion of the cell. Conversely, if the
backness pathway is suppressed, the frontness pathways invade and
occupy the previous back of the cell.
\item Depending on the parameter values, the model yields steady states
corresponding to both chemoattraction and chemorepulsion. Analysis
of the model suggests that chemorepulsion-to-chemoattraction transitions
observed in neurons can be triggered by agents that inhibit the backness
pathways more than the frontness pathways.
\end{enumerate}
The model provides a useful starting point for formulating models
that account for the positive and negative feedback effects, both
of which have been shown to play a role in gradient sensing.

\subsubsection*{Acknowledgments}

I would like to thank Prof.~D. A. Lauffenburger for helpful comments
on the manuscript.

\bibliographystyle{unsrt}
\bibliography{C:/lyx/Bibfiles/sensing,C:/lyx/Bibfiles/sensing_karthik,C:/lyx/Bibfiles/growthKinetics,C:/lyx/Bibfiles/neuron}

\appendix

\section{\label{a:HSSstability}Existence/stability of homogeneous steady
states}

The homogeneous steady states of (\ref{eq:modelDeq1}--\ref{eq:modelDeq3})
satisfy the equations\begin{align*}
-u_{1}+a_{12}u_{2}+a_{13}u_{3} & =0\\
\left(r-a_{21}u_{1}-a_{22}u_{2}-a_{23}u_{3}\right)u_{2} & =0\\
\left(r-a_{31}u_{1}-a_{32}u_{2}-a_{33}u_{3}\right)u_{3} & =0.\end{align*}
It follows that there are 4 possible steady states: $u_{1}=u_{2}=u_{3}=0$,
denoted $E_{000}$; $u_{1},u_{2}>0,u_{3}=0$, denoted $E_{110}$;
$u_{1},u_{3}>0,u_{2}=0$, denoted $E_{101}$; and $u_{1},u_{2},u_{3}>0$,
denoted $E_{111}$. In what follows, we show the existence and stability
criteria for all 4 steady states. To this end, it is convenient to
note that the Jacobian at any point, $u_{1},u_{2},u_{3}$ is\[
J=\left[\begin{array}{ccc}
-1 & a_{12} & a_{13}\\
-\rho_{2}u_{2}a_{21} & -\rho_{2}u_{2}a_{22}+\rho_{2}f_{2} & -\rho_{2}u_{2}a_{31}\\
-\rho_{3}u_{3}a_{31} & -\rho_{3}u_{3}a_{32} & -\rho_{3}u_{3}a_{33}+\rho_{3}f_{3}\end{array}\right]\]
where\begin{align*}
f_{2} & \equiv r-a_{21}u_{1}-a_{22}u_{2}-a_{23}u_{3}\\
f_{3} & \equiv r-a_{31}u_{1}-a_{32}u_{2}-a_{33}u_{3}\end{align*}

\begin{figure}
\begin{center}\subfigure[]{\includegraphics[%
  width=3in,
  height=2in]{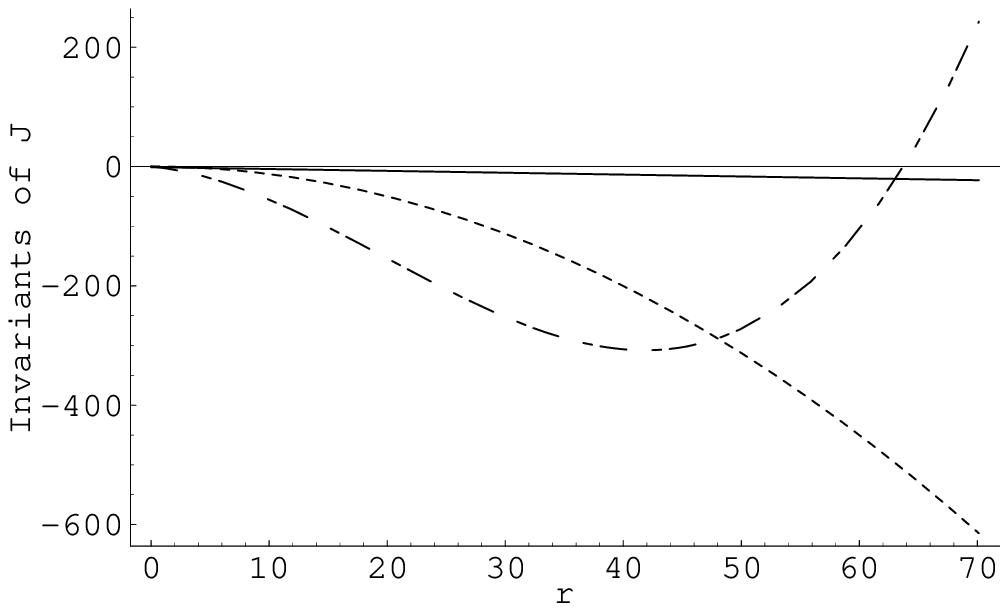}}\hspace{0.1in}\subfigure[]{\includegraphics[%
  width=3in,
  height=2in]{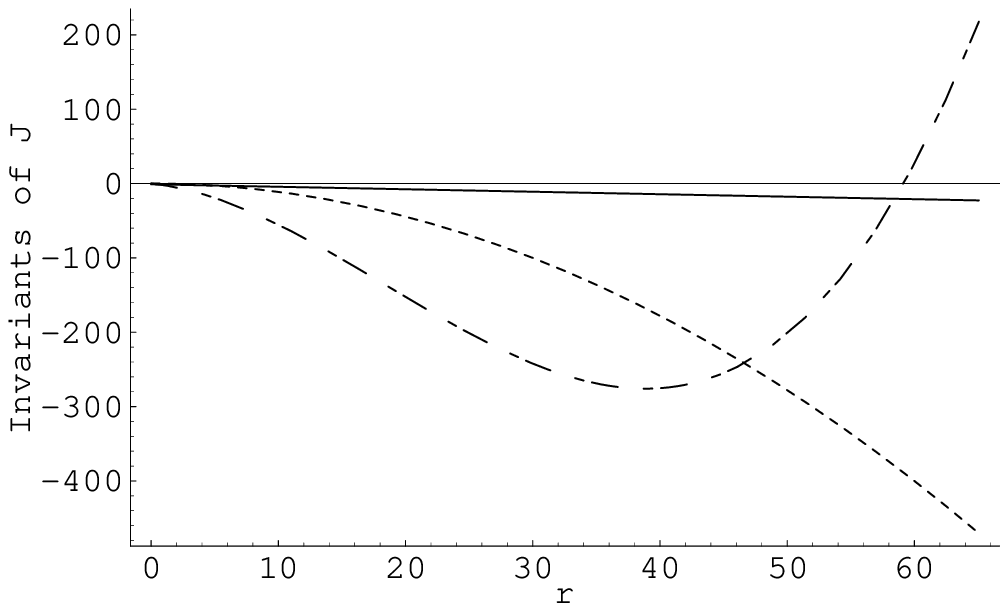}}\end{center}

\caption{\label{f:Invariants}Variation of $\textnormal{tr }J$ (---), $\det J$
(-- --), and $(\textnormal{tr }J)(\Sigma J)-\det J$ (--- ---) as
a function of $r$. (a)~ Parameter values chosen such that $U_{2}$
prevails over $U_{3}$ in the absence of $U_{1}$. (b) ~ Parameter
values chosen such that $U_{3}$ prevails over $U_{2}$ in the absence
of $U_{1}$. }
\end{figure}

\begin{enumerate}
\item $E_{000}=[0,0,0]^{t}$: It is evident that this steady state always
exists. The Jacobian yields no information regarding the stability
since it is singular at $E_{000}$. However, we can infer that it
is unstable for all $r>0$ by observing that the $u_{2}$-axis is
an invariant manifold. The motion along this invariant manifold is
given by \[
\frac{du_{2}}{dt}=\rho_{2}(r-a_{22}u_{2})u_{2},\]
which drives the system away from $E_{000}$ for all $r>0$.
\item $E_{110}=[u_{1},u_{2},0]^{t},\; u_{1},u_{2}>0$: It is easy to check
that this steady state exists for all $r>0$, and is given by\[
u_{3}=0,\; u_{2}=\frac{r}{\alpha_{22}},\; u_{1}=a_{12}u_{2},\]
where $\alpha_{22}\equiv a_{22}+a_{21}a_{12}$, and the Jacobian at
$E_{110}$ is\begin{equation}
\left[\begin{array}{ccc}
-1 & a_{12} & a_{13}\\
-\rho_{2}u_{2}a_{21} & -\rho_{2}u_{2}a_{22} & -\rho_{2}u_{2}a_{31}\\
0 & 0 & \rho_{3}(r-a_{31}u_{1}-a_{32}u_{2})\end{array}\right].\label{eq:JacobianE110}\end{equation}
It follows that one of the eigenvalues is \[
\lambda_{1}=\rho_{3}(r-a_{31}u_{1}-a_{32}u_{2})=\rho_{3}r\left(1-\frac{\alpha_{32}}{\alpha_{33}}\right),\;\]
where $\alpha_{32}\equiv a_{32}+a_{31}a_{12}$. The other two eigenvalues
satisfy\begin{alignat*}{1}
\lambda_{2}+\lambda_{3} & =-1-\rho_{2}a_{22}ru_{2}<0,\\
\lambda_{2}\lambda_{3} & =\rho_{2}a_{22}ru_{2}+\rho_{2}u_{2}a_{21}a_{12}>0,\end{alignat*}
so that the real parts of $\lambda_{2},\lambda_{3}$ are negative.
Hence, $E_{110}$ is stable if and only if $\beta_{32}\equiv\alpha_{32}/\alpha_{22}>1$.
\item $E_{101}=[u_{1},0,u_{3}]^{t},\; u_{1},u_{3}>0$: The calculations
are analogous to those for $E_{101}$.
\item $E_{111}=[u_{1},u_{2},u_{3}]^{t},\; u_{1},u_{2},u_{3}>0$: This steady
state satisfies the equations\begin{align*}
u_{1} & =a_{12}u_{2}+a_{13}u_{3}\\
r & =\alpha_{22}u_{2}+\alpha_{23}u_{3}\\
r & =\alpha_{32}u_{2}+\alpha_{33}u_{3}.\end{align*}
It follows that $E_{111}$ exists for all $r>0$ if and only if $[1,1]^{t}$
lies between $[\alpha_{22},\alpha_{32}]^{t}$ and $[\alpha_{23},\alpha_{33}]^{t}$,
i.e., $\beta_{23}\equiv\alpha_{23}/\alpha_{33},\beta_{32}\equiv\alpha_{32}/\alpha_{22}$
satisfy\[
\beta_{23},\beta_{32}<1\textnormal{ or }\beta_{23},\beta_{32}>1.\]
In both cases, the steady state is given by\begin{align*}
u_{2} & =\frac{1}{\alpha_{22}}\frac{1-\beta_{23}}{1-\beta_{23}\beta_{32}}r,\\
u_{3} & =\frac{1}{\alpha_{33}}\frac{1-\beta_{32}}{1-\beta_{23}\beta_{32}}r,\\
u_{1} & =a_{12}u_{2}+a_{13}u_{3}.\end{align*}
Evidently, $u_{1},u_{2},u_{3}$ increase linearly with $r$.\\
It turns out that $E_{111}$ is unstable if $\beta_{23},\beta_{32}>1$,
and stable for sufficiently small $r>0$ if $\beta_{23},\beta_{32}>1$.
To see this, observe that the Jacobian at $E_{111}$ is\begin{equation}
\left[\begin{array}{ccc}
-1 & a_{12} & a_{13}\\
-\rho_{2}u_{2}a_{21} & -\rho_{2}u_{2}a_{22} & -\rho_{2}u_{2}a_{31}\\
-\rho_{3}u_{3}a_{31} & -\rho_{3}u_{3}a_{32} & -\rho_{3}u_{3}a_{33}\end{array}\right],\label{eq:JacobianE111}\end{equation}
so that \begin{alignat*}{1}
\textnormal{tr }J & =-1-\rho_{2}a_{22}u_{2}-\rho_{3}a_{33}u_{3},\\
\det J & =-\rho_{2}\rho_{3}u_{2}u_{3}\alpha_{22}\alpha_{23}(1-\beta_{23}\beta_{32}),\\
\Sigma\, J & =\rho_{2}\rho_{3}u_{2}u_{3}(a_{22}a_{33}-a_{23}a_{32})+\rho_{2}u_{2}\alpha_{22}\\
 & \qquad+\rho_{3}u_{3}\alpha_{33}.\end{alignat*}
If $\beta_{23},\beta_{32}>1$, then $\det J>0$, and $E_{111}$ is
unstable. Hence, $E_{111}$ is stable only if $\beta_{23},\beta_{32}<1$.
We are particularly interested in the case when $a_{22}a_{33}-a_{23}a_{32}>0$,
since this is necessary for Turing instability (see below). Under
these conditions, $E_{111}$ is stable for all sufficiently small
$r>0$, since $\textnormal{tr }J<0$, $\det J<0$ for all $r>0$,
and $(\textnormal{tr }J)(\Sigma J)-\det J<0$ for all sufficiently
small $r>0$ (Figure~\ref{f:Invariants}).
\end{enumerate}

\section{\label{a:HSSturingStability}Turing instability of homogeneous steady
states}

\subsection{Conditions for Turing instability}

A homogeneous steady state, $u=[u_{1}(r),u_{2}(r),u_{3}(r)]^{t}$,
is Turing unstable if it is stable in the absence of diffusion and
unstable in the presence of diffusion. Now the stability of the homogeneous
steady state, $u$, in the presence of diffusion is determined by
the linearized equation\begin{equation}
\frac{\partial v}{\partial t}=D\frac{\partial v}{\partial x^{2}}+J(u)v\label{eq:LinearizedRDeqn}\end{equation}
where $v$ denotes the deviation from $u$, $J(u)$ denotes the Jacobian
at $u$, and $D=\textnormal{diag}\mathnormal{(d_{1},d_{2},d_{3})}$.
Let $\mu_{k}$ and $\phi_{k}(x)$ denote the eigenvalues and eigenfunctions
of the differential operator, $\frac{d^{2}}{dx^{2}}$, with Neumann
boundary conditions, i.e.,\[
\mu_{k}=-(k\pi)^{2},\;\phi_{k}(x)=\cos(k\pi x),\; k=0,1,2,\ldots\]
If the steady state is perturbed along any eigenfunction, $\phi_{k}(x)$,
the subsequent evolution of the perturbation is of the form $v(t,x)=s_{k}(t)\phi_{k}(x)$
where $s_{k}(t)\in\mathbb{R}^{3}$. Substituting in (\ref{eq:LinearizedRDeqn})
yields \[
\frac{ds_{k}}{dt}=C_{k}s_{k},\; C_{k}(u)\equiv J(u)-(k\pi)^{2}D.\]
It follows that the homogeneous steady state is stable if and only
if the eigenvalues of $C_{k}(u)$ have negative real parts for all
$k\ge0$. It is Turing unstable if and only if

\begin{figure}
\begin{center}\subfigure[]{\includegraphics[%
  width=3in,
  height=2in]{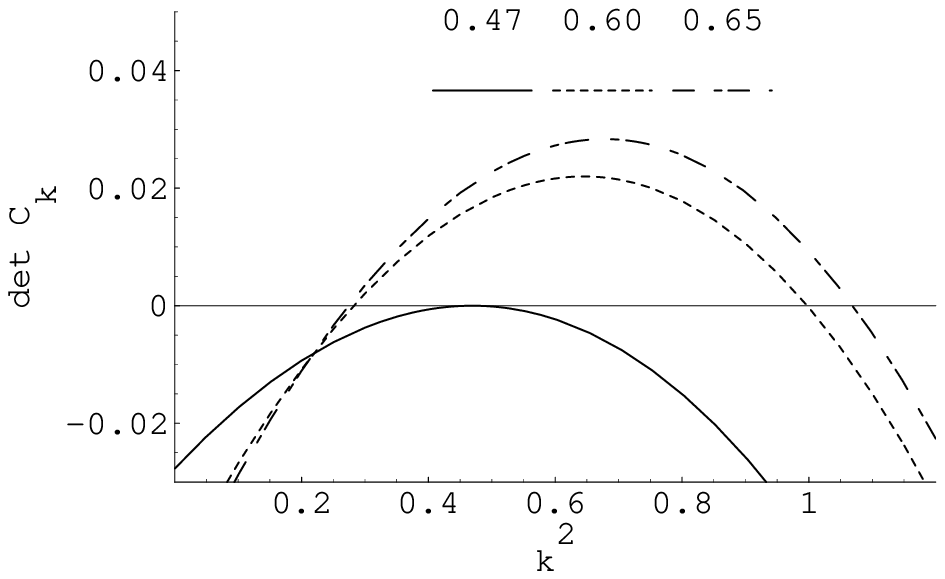}}\hspace{0.1in}\subfigure[]{\includegraphics[%
  width=3in,
  height=2in]{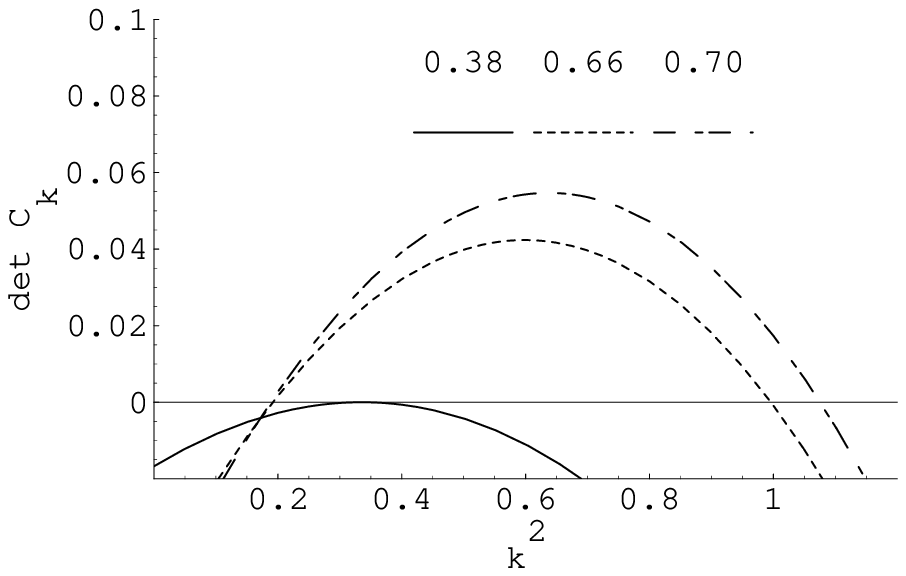}}\end{center}

\caption{\label{f:detCk}Variation of $\det C_{k}$ as a function of $k^{2}$
at various values of $r$. (a)~ Parameter values chosen such that
$U_{2}$ prevails over $U_{3}$ in the absence of $U_{1}$. (b) ~
Parameter values chosen such that $U_{3}$ prevails over $U_{2}$
in the absence of $U_{1}$. }
\end{figure}

\begin{enumerate}
\item The eigenvalues of $C_{0}(u)=J(u)$ have negative real parts (which
ensures that the homogeneous steady state is stable in the absence
of diffusion).
\item The eigenvalues of $C_{k}(u)$ have positive real part for some $k>0$
(which ensures that the homogeneous steady state is unstable in the
presence of diffusion).
\end{enumerate}
The Turing bifurcation occurs at the critical value, $r=r_{0}>0$,
such that exactly one of the eigenvalues of $C_{k}(u)$ becomes zero,
i.e.,\[
\det C_{k}(u)=0,\frac{d}{dk^{2}}\det C_{k}(u)=0\]
for some $k>0$ (Figure~\ref{f:detCk}).

\subsection{Turing instability of the homogeneous steady states }

It follows from the definition of $C_{k}$ that~\cite{Satnoianu2000}\begin{align}
\det C_{k}(u) & =\triangle_{123}-(d_{1}\triangle_{23}+d_{2}\triangle_{13}+d_{3}\triangle_{13})k^{2}\nonumber \\
 & \qquad+(d_{1}d_{2}\triangle_{3}+d_{2}d_{3}\triangle_{1}+d_{1}d_{3}\triangle_{2})k^{4}\nonumber \\
 & \qquad-(d_{1}d_{2}d_{3})k^{6}\label{eq:detCk}\end{align}
where $\triangle_{i}$, $\triangle_{ij},i\ne j$, $\triangle_{123}$
are the determinants of the 1-, 2-, and 3-dimensional subsystems of
$J$, i.e., $\triangle_{i}=J_{ii}(u)$, \begin{alignat*}{1}
\triangle_{ij} & =\det\left[\begin{array}{cc}
J_{ii} & J_{ij}\\
J_{ji} & J_{jj}\end{array}\right],\; i\ne j\end{alignat*}
 and $\triangle_{123}=\det J(u)$. These results are sufficient for
investigating the possibility that the homogeneous steady states undergo
a Turing bifurcation.

We consider each of the homogeneous steady states.

\begin{enumerate}
\item $E_{000}=[0,0,0]^{t}$: Since $E_{000}$ is unstable even in the absence
of diffusion, it cannot entertain Turing instability.
\item $E_{110}=[u_{1},u_{2},0]^{t}$: This steady state is stable in the
absence of diffusion provided $\beta_{32}>1$. Thus, it can undergo
a Turing instability if and only if there is an $r>0$ such that (\ref{eq:detCk})
has a positive root. It turns out that such a root does not exist.
To see this, observe that since $E_{110}$ is stable in the absence
of diffusion, $\triangle_{123}=\det J<0$. Inspection of the Jacobian
(\ref{eq:JacobianE110}) shows that the $\triangle_{i}$'s are always
negative, and the $\triangle_{ij}$'s are always positive. Hence,
all the coefficients of the polynomial (\ref{eq:detCk}) are negative
for all $r>0$. Such a polynomial cannot have positive roots ($k>0$).
We conclude that $E_{110}$ cannot undergo a Turing instability.
\item $E_{101}=[u_{1},0,u_{3}]^{t}$: This steady state is stable in the
absence of diffusion provided $\beta_{23}<1$. An argument analogous
to that for $E_{110}$ shows that it cannot undergo a Turing instability.
\item $E_{111}=[u_{1},u_{2},u_{3}]^{t}$: This steady state is stable in
the absence of diffusion provided $\beta_{23},\beta_{32}<1$ and $r$
is sufficiently small. Under these conditions, $\triangle_{123}<0$.
Inspection of the Jacobian (\ref{eq:JacobianE111}) shows that no
matter what the value of $r$, all the $\triangle_{i}$'s are negative,
and $\triangle_{12},\triangle_{13}$ are positive. Hence, $E_{111}$
can undergo a Turing instability only if $\triangle_{23}<0$. This
condition is not only necessary, but also sufficient. For, if $d_{2}$
and $d_{3}$ are sufficiently small compared to $d_{1}$, the coefficient
of $k^{2}$ is approximately $d_{1}\triangle_{23}<0$, and the polynomial
(\ref{eq:detCk}) has a positive root.\\
The non-homogeneous steady state that bifurcates from $E_{111}$ is
such that $U_{2}$ and $U_{3}$ are out-of-phase. To see this, it
suffices to observe that to a first degree of approximation, the non-homogeneous
steady state near the Turing bifurcation point has the form $v_{0}\cos(k_{0}x)$,
where $k_{0}$ be the critical wavenumber, and $v_{0}$ is the eigenvector
corresponding to the zero eigenvalue of $C_{k_{0}}$, i.e., $C_{k_{0}}v_{0}=0$.
The null space of the rank-2 matrix, $C_{k_{0}}$, is spanned by the
vector, $[c_{1},c_{2},1]$, where \begin{align*}
c_{2} & \equiv-\frac{C_{k_{0},22}-C_{k_{0},12}C_{k_{0},21}/C_{k_{0},11}}{C_{k_{0},23}-C_{k_{0},21}C_{k_{0},13}/C_{k_{0},11}}.\end{align*}
Substitution of the coefficients of $C_{k_{0}}$ shows that $c_{2}$
is always negative, which implies that $U_{2}$ and $U_{3}$ are out-of-phase.\end{enumerate}

\end{document}